\documentclass[fleqn,usenatbib]{mnras}

\usepackage{newtxtext,newtxmath}
\usepackage{xcolor}

\usepackage[T1]{fontenc}

\DeclareRobustCommand{\VAN}[3]{#2}
\let\VANthebibliography\thebibliography
\def\thebibliography{\DeclareRobustCommand{\VAN}[3]{##3}\VANthebibliography}

\usepackage{graphicx}	
\usepackage{amsmath}	
\DeclareUnicodeCharacter{2212}{-}

\newcommand{\CIV}{\ion{C}{iv}}
\newcommand{\kms}{km\,s$^{-1}$}

\usepackage{soul}
\newcommand{\hli}[1]{#1}

\defcitealias{rankine_2020}{R20}
\defcitealias{gibson_catalog_2009}{G09}
\defcitealias{wu_x-ray_2010}{W10}

\usepackage{orcidlink}

\title[X-ray selected BAL quasars]{X-ray selected broad absorption line quasars in SDSS-V: BALs and non-BALs span the same range of X-ray properties}

\author[P. Hiremath et al.]{Pranavi Hiremath$^{\orcidlink{0009-0005-0072-6973}}$,$^{1}$
Amy L. Rankine$^{\orcidlink{0000-0002-2091-1966}}$,$^{1}$\thanks{E-mail: amy.rankine@ed.ac.uk (ALR)}
James Aird$^{\orcidlink{0000-0003-1908-8463}}$,$^{1}$
W. N. Brandt$^{\orcidlink{0000-0002-0167-2453}}$,$^{2, 3, 4}$
Paola Rodr\'iguez Hidalgo$^{\orcidlink{0000-0003-0677-785X}}$,$^{5}$
\newauthor
Scott F. Anderson$^{\orcidlink{0000-0002-6404-9562}}$,$^{6}$
Catarina Aydar$^{\orcidlink{0000-0001-5609-2774}}$,$^{7, 8}$
Claudio Ricci$^{\orcidlink{0000-0001-5231-2645}}$,$^{9, 10}$
Donald P. Schneider$^{\orcidlink{0000-0001-7240-7449}}$,$^{2, 3}$
M. Vivek$^{11, 12}$
\newauthor
Zsofi Igo$^{\orcidlink{0000-0001-9274-1145}}$,$^{7, 8}$
Sean Morrison$^{\orcidlink{0000-0002-6770-2627}}$,$^{13}$
Mara Salvato$^{\orcidlink{0000-0001-7116-9303}}$$^{7}$
\\
$^{1}$Institute for Astronomy, University of Edinburgh, Royal Observatory, Blackford Hill, Edinburgh EH9 3HJ, UK\\
$^{2}$Department of Astronomy and Astrophysics, The Pennsylvania State University, University Park, PA 16802, USA\\ 
$^{3}$Institute for Gravitation and the Cosmos, The Pennsylvania State University, University Park, PA 16802, USA\\ 
$^{4}$Department of Physics, 104 Davey Laboratory, The Pennsylvania State University, University Park, PA 16802, USA\\ 
$^{5}$Physical Sciences Division, School of STEM, University of Washington Bothell, Bothell WA, 98011, USA\\ 
$^{6}$Department of Astronomy, University of Washington, Box 351580, Seattle, WA 98195, USA\\ 
$^{7}$Max Planck Institute for Extraterrestrial Physics, Giessenbachstrasse 1, 85748 Garching, Germany\\ 
$^{8}$Excellence Cluster ORIGINS, Boltzmannstrasse 2, D-85748 Garching, Germany\\ 
$^{9}$Instituto de Estudios Astrof\'isicos, Facultad de Ingenier\'ia y Ciencias, Universidad Diego Portales, Av. Ej\'ercito Libertador 441, Santiago, Chile\\ 
$^{10}$Kavli Institute for Astronomy and Astrophysics, Peking University, Beijing 100871, China\\ 
$^{11}$Indian Institute of Astrophysics, Koramangala II block, Bangalore - 560034, Karnataka, India\\ 
$^{12}$Astronomisches Rechen-Institut, Zentrum fur Astronomie der Universitat Heidelberg, Monchhofstr. 12-14, D-69120 Heidelberg, Germany\\ 
$^{13}$Department of Astronomy, University of Illinois at Urbana-Champaign, Urbana, IL 61801, USA\\ 
}

\date{Accepted XXX. Received YYY; in original form ZZZ}

\pubyear{\the\year{}}

\begin{document}
\label{firstpage}
\pagerange{\pageref{firstpage}--\pageref{lastpage}}
\maketitle

\begin{abstract}
Broad absorption line (BAL) quasars are often considered X-ray weak relative to their optical/UV luminosity, whether intrinsically (i.e., the coronal emission is fainter) or due to large column densities of absorbing material. The SDSS-V is providing optical spectroscopy for samples of quasar candidates identified by eROSITA as well as \textit{Chandra}, \textit{XMM} or \textit{Swift}, making the resulting datasets ideal for characterising the BAL quasar population within an X-ray selected sample. We use the Balnicity Index (BI) to identify the BAL quasars based on absorption of the {\CIV}$\lambda\,1549$ emission line in the optical spectra, finding 143 BAL quasars in our sample of 2317 X-ray selected quasars within $1.5\le z \le3.5$. This observed BAL fraction of $\approx$\,6 per cent is comparable to that found in optically selected samples. We also identify absorption systems via the Absorption Index (AI) which includes mini-BALs and NALs, finding 954 quasars with AI $>0$. We consider the {\CIV} emission space (equivalent width vs. blueshift) to study the BAL outflows within the context of the radiatively driven accretion disc-wind model. X-ray selection excludes the highest outflow velocities in emission but includes the full range of absorption velocities which we suggest is consistent with the BAL gas being located further from the X-ray corona than the emitting gas. We observe both X-ray weak and X-ray strong BALs (via the optical-to-X-ray spectral slope, $\alpha_\text{ox}$) and detect little evidence for differing column densities between the BAL and non-BAL quasars, suggesting the BALs and non-BALs have the same shielding gas and intrinsic X-ray emission.

\end{abstract}




\section{Introduction}\label{sec:intro}

At the centre of every massive galaxy, a common consensus is the existence of a supermassive black hole (SMBH) that goes through active phases during its lifetime, accreting material from the inner regions of the galaxy and is therefore known as an active galactic nucleus \citep[AGN;][]{Lynden-Bell1969}. The varying scales in the model of an AGN (see \citealt{Heckman2014}), along with the observational evidence found for the important role AGN play in the formation and evolution of galaxies (e.g., \citealt{Magorrian_1998}; \citealt{aird}; \citealt{brandt_surveys_2022}), suggest the presence of feedback. The spectral energy distribution (SED) of an AGN covers the full electromagnetic spectrum meaning that AGN populations can be identified at X-ray, ultra-violet (UV), optical, as well as other wavelengths. 

Quasars are the brightest of all AGN and represent the most radiatively efficient phase of SMBH growth, releasing energy as kinetic luminosity through sub-relativistic wide-angle winds \citep[e.g.][]{Silk1998}. Broad absorption line (BAL) quasars are a subset of quasars with rest-frame UV spectra that show evidence for outflows via the presence of blue-shifted absorption lines, usually identified based on absorption blueward of the {\CIV}\,$\lambda$1549 broad emission line. The definition most frequently used to define a BAL quasar requires absorption troughs with widths $\geq2000$\,{\kms} \citep{Weymann1991}. Other definitions can include much narrower absorption features \citep[$\ge$ 450\,{\kms};][]{Hall_2002} often called mini-BALs; however, this can result in contamination by strong narrow absorption lines \citep[NALs: FWHM $<500$\,\kms; e.g.,][]{hamann_elemental_1999, Bowler2014}.
Quasars can additionally have absorption in lower-ionization species such as \ion{Al}{iii}\,$\lambda$1857 and \ion{Mg}{ii}\,$\lambda$2800 -- the so-called LoBALs -- however we will focus on BAL quasars with absorption only in the high-ionization lines -- the HiBALs, hereafter referred to as BALs. While the driver of BAL quasar outflows may be radiation, thermal, or magnetic pressure, BAL features are strong evidence for significant momentum transfer from a powerful radiation field to the gas, which results in the propulsion of the gas to high velocities (e.g., \citealt{Weymann1991}; \citealt{allen}). These observations support a radiatively driven hydrodynamical scenario for quasar outflows whereby outflowing winds are propelled by radiation pressure caused by the accretion disc's UV emissions \citep{Proga2007}. Correlations between outflow velocity, Eddington ratio, and luminosity lead \citet{vivek_broad_2025} to conclude that radiation pressure is important for driving outflows, but suggest that additional mechanisms (i.e., greater outflow distances, thicker discs, softer SEDs) are required to explain the presence of BALs in low-Eddington ratio sources.

\citet{Richards2011} demonstrated that the distribution of radio-quiet/loud quasars in the {\CIV} emission space ({\CIV} equivalent width [EW] vs. blueshift) can be well understood via a `disc' vs. `wind' model. In detail, a disc-dominated quasar would be expected to have symmetric line profiles (low blueshift) with the line dominated by the disc emission, and a wind-dominated system would have lower {\CIV} EW and high blueshift. 
\citet{rivera_2020} and \citet{richards_2021} defined the `{\CIV} distance' metric in order to group quasars along the best-fitting line through {\CIV} space with higher {\CIV} distance values associated with stronger winds. \ion{He}{ii} EW (an SED indicator; \citealt{Leighly2004}) is known to correlate with the {\CIV} emission space (\citealt{Baskin_2013, Baskin2015-xk}; \citealt{rankine_2020} [hereafter \citetalias{rankine_2020}]; \citealt{Timlin2021}) and \citet{Rivera_2022} showed that \ion{He}{ii} EW is correlated most strongly with {\CIV} distance over blueshift or EW. Thus the {\CIV} distance is a convenient way to explore the impact of changes in the SED under a disc-wind model. Such an accretion disc-wind model is often invoked to explain observed BAL features (see figure 1 of \citealt{Luo2013}) and according to this model, radiatively driven winds launched from the accretion disc result in BALs when inclination angles are large, implying the wind is in the direct line-of-sight \citep{Leighly2015}. Here, these accretion disc winds may be equatorial \citep[e.g.][]{Murray1995} and/or polar \citep[e.g.][]{punsly_1999}, and can be continuous or consist of knots/blobs of matter (\citealt{ghisellini_2004}; \citealt{matthews_testing_2016}). Therefore, BAL quasars can be considered generally to be wind-dominated quasars which were investigated by \citetalias{rankine_2020} within the {\CIV} space. They reported that the optically selected BAL quasars have moderate EW and high blueshift but crucially, however, they can be found in all regions of the space. Additionally, \citet{ahmed_exploring_2025} found that BALs and non-BALs have the same SEDs, from mid-infrared to X-ray. 

Typically 10-20 per cent of optically selected quasars show broad absorption features; however, the intrinsic fraction is expected to be higher (up to 40 per cent; \citealt{allen}; \citealt{dai_2012}) due to considerations of geometrical and evolutionary aspects of BAL quasar outflows. Specifically, in a geometrical scenario, the fraction of BALs observed depends on the solid angle the BAL winds subtend \citep[e.g.][]{ghosh_2007} while an evolutionary scenario considers the variability in the appearance/disappearance of the BAL winds \citep[e.g.][]{mishra_2021}. Regardless of the scenario, it implies that a classified non-BAL quasar could be harboring BAL winds that are not currently within the line-of-sight and therefore making the intrinsic fraction of BAL quasars difficult to ascertain. Moreover, the observed fraction of BAL quasars also depends on the signal-to-noise (S/N) ratio of the sample studied 
(\citealt{gibson_catalog_2009} [hereafter \citetalias{gibson_catalog_2009}]; \citealt{allen}).

Additionally, to ensure the outflowing gas is not over-ionized by the X-ray and extreme-UV radiation from the corona/accretion disc i.e., to explain the observed high velocities, the disc-wind model also invokes X-ray shielding gas with high column densities \citep{Murray1995}. This gas allows for efficient line driving of the UV disc winds and is a natural consequence in the hydrodynamical 2D simulations of \citet{proga_dynamics_2004}. \citet{Luo2013} considered the X-ray shielding material to originate from `failed winds' i.e., gas that was unable to reach the local escape velocity due to overionization. \citet{baskin_radiation_2014} provide an alternate theory via Radiation Pressure Compression or Confinement (RPC) whereby
radiation pressure compresses the gas along the radial direction, resulting in
a high electron density and low ionization parameter
to allow absorption in the UV
and efficient driving of the outflows to high velocities. 
This theory does not invoke X-ray shielding gas but rather suggests the radiation force itself can lead to high column densities through compression of the absorbing gas, preventing it from becoming over-ionized. Many studies have indeed shown that BAL quasars often appear to be X-ray weak relative to non-BAL quasars (e.g., \citealt{green_1995}; \citealt{laor_1997}; \citealt{brandt_2000}; \citealt{gallagher_exploratory_2006};  \citetalias{gibson_catalog_2009}; \citealt{Luo2014}; \citealt{saccheo2023}). 

The X-ray weakness of BAL quasars can be studied by quantifying the relationship between X-ray and UV luminosities. Studies such as \citet{Avni1982} parameterized this relationship between X-ray ($L_\textup{X}$) and UV luminosity ($L_\textup{UV}$) as $L_\textup{X}$ $\propto$ $L_\textup{UV}^{\gamma}$ with $\gamma$ $\sim$ 0.6. The relationship between $L_\textup{X}$-$L_\textup{UV}$ is often quantified using $\alpha_\text{ox}$, which is the energy index or slope that connects the optical with the X-ray band for a power law spectrum ($\textup{F}_{\nu}$ $ \propto$ $\nu^{\alpha}$). In detail, \citet{Tananbaum1979} introduced this correlation specifically between monochromatic luminosities at 2500\,\AA\ and 2\,keV, although the physical mechanism responsible for this relation remains unclear, i.e., the relationship between the UV disc and the X-ray emissions from the hot corona. \hli{Moreover, the relationship between $\alpha_\text{ox}$ and $L_{2500}$ is often observed to be an anti-correlation }\citep[e.g.,][]{steffen_x-ray--optical_2006, lusso_tight_2016, Rankine2023},\hli{ which implies that as $L_{2500}$ increases the corona becomes (relatively) weaker and thus the AGN becomes more disc-dominated. However, there is intrinsic scatter in this relation and objects with greater than average X-ray luminosities given their optical luminosity are expected to drive winds less efficiently} \citep{Giustini2019, Timlin2021, Rivera_2022, temple_testing_2023}.
While the choice of $L_{2500}$ is somewhat arbitrary, recent studies such as \citet{Timlin2021} and \citet{Jin2023} have shown that the $L_{2500}$ alone is sufficient to describe the UV/optical emissions of a quasar over a broader wavelength range, making $L_{2500}$ suitable for investigating the relationship between UV and X-ray emissions.

\citet{laor_1997} demonstrated that the X-ray flux of BAL quasars was 10-30 times lower than expected for their UV flux, classifying them as ``soft-X-ray weak'' objects and \citet{laor_2002} found that these quasars had the strongest absorption features. \citetalias{gibson_catalog_2009} (further studied in \citealt{wu_x-ray_2010}; \citetalias{wu_x-ray_2010} hereafter) also found that observed X-ray weakness was correlated to maximum velocity and absorption strengths of the {\CIV} BAL troughs, suggesting the impact of X-ray absorption on UV disc winds and further supporting the accretion disc wind model.  
However, \citet{Giustini2008} studied a sample of 54 X-ray-selected quasars to find that one-third of the sample showed little to no absorption and the distributions of the optical-to-X-ray spectral slope, $\alpha_{\text{ox}}$, of the BAL and non-BAL populations were similar, suggesting that these BAL quasars were not X-ray weaker relative to the non-BALs. They attributed their results to the X-ray selected sample preferentially being viewed from smaller angles with respect to the accretion disc rotation axis than optically selected BAL quasars. While they conducted X-ray spectral analysis considering a neutral absorber model, \citet{streblyanska_2010} considered both neutral and ionized absorption models for a sample of 39 X-ray-selected BAL quasars, 
finding that while 36 per cent of BALs did not show absorption in a neutral absorption model, 90 per cent of them were absorbed in a physically motivated ionized model, although even this model is likely to be too simplistic.

This paper aims to study the multiwavelength properties of X-ray selected (including eROSITA, \textit{Chandra}, \textit{XMM}, and \textit{Swift}) SDSS-V BAL quasars relative to the non-BAL quasars to further investigate the X-ray weakness of ``classical'' BAL quasars previously predominantly observed in optically selected samples. This observed X-ray weakness of BAL quasars makes them interesting candidates to study within the X-ray-selected SDSS-V sample, so the preliminary aim of this paper is to find the fraction of BAL quasars present in this sample. The eROSITA/SDSS-V collaboration marks the first time we have a large (soft) X-ray-selected sample with optical and corresponding X-ray data readily available. Therefore, we can build on the previous investigations of X-ray-selected BAL quasars through statistical analyses of a larger sample. Additionally, the work in this paper can be considered to be an extension of \citetalias{rankine_2020}; while they focused on the study of emission and absorption properties of optically selected BAL quasars relative to non-BAL quasars, we aim to extend this investigation to X-ray-selected BAL quasars to obtain a more complete picture for the BAL population. The sample used and the results found by \citetalias{rankine_2020} form a basis of comparison to better understand the X-ray-selected sample studied in this paper. 

The outline of this paper is as follows: we first describe in further detail our sample in section \ref{sec:data}. Section \ref{sec:BAL identification} presents the BAL identification methods employed to achieve the first aim of this paper, i.e., identify X-ray-selected BAL quasars. Section \ref{sec:CIVspace} investigates the identified sub-populations by placing them in the {\CIV} emission space and further explores the identified BAL quasars in detail. Section \ref{sec:X-ray prop} studies the X-ray properties of the BAL and non-BAL quasars by determining their X-ray luminosities, $\alpha_\text{ox}$ and $\Delta\alpha_\text{ox}$ values. Section \ref{sec:discussion} discusses and section \ref{sec:conclusions} summarizes the obtained results.

A $\Lambda$CDM cosmology with $h_0$ = 0.71, $\Omega_\textup{M}$ = 0.27, and $\Omega_{\Lambda}$ = 0.73 is adopted for determining quantities such as quasar luminosities.

\section{Data}\label{sec:data}

\begin{figure}
	\includegraphics[width=\columnwidth]{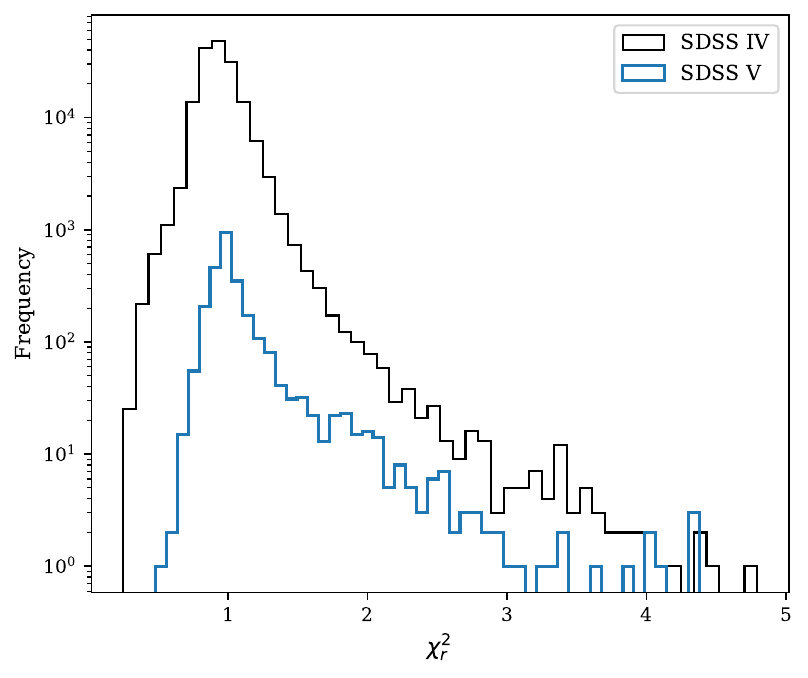}
    \caption{Distribution of the reduced ${\chi}^2$ values of the SDSS-IV and SDSS-V samples consisting of $\sim$165,000 and 2680 quasars respectively, show that the two samples are comparable with a common peak occurring at ${\chi}_r^2$ $\approx$ 1. This suggests that the MFICA components generated using SDSS-IV are suitable for constructing a smooth, high S/N continuum for the spectra in the SDSS-V sample.} 
    \label{fig:chisqr}
\end{figure}

The eROSITA (extended ROentgen Survey with an Imaging Telescope Array) instrument aboard the SRG \citep[Russian–German Spektrum Roentgen Gamma;][]{sunyaev_srg_2021} space mission, launched in July 2019, presents a step forward for the exploration of X-ray AGN properties \citep{Predehl}. eROSITA is an X-ray-sensitive telescope capable of discovering new X-ray sources within the energy range of $\sim$0.2--8\,keV \citep{Merloni}. SDSS-V sets out to be the first of its kind to provide multi-epoch, all-sky surveys in the optical and IR wavelengths \citep{kollmeier_sloan_2025}. The black hole mapper program (Anderson et al. in prep) of SDSS-V forms one of the three top-level goals aimed to provide spectroscopic measurements such as redshifts for $\sim$300,000 X-ray sources identified by eROSITA. 

Optical spectra have been obtained using the BOSS multi-fibre spectrographs \citep{smee_2013} on the Apache Point Observatory 2.5\,m SDSS telescope \citep{Gunn_2006} and the 2.54\,m Irénée du Pont Telescope at Las Campanas Observatory \citep{bv73}. \hli{We select sources with redshifts in the range 1.5--3.5 and S/N $>$ 5 per SDSS spectrum pixel ($\Delta v = 69$\,\kms) within the wavelength range of investigation (1260-3000 \AA), and spectroscopically classified as ``QSO'' by the SDSS pipeline.} Our sample consists of 2680 objects targeted by SDSS-V \citep{kollmeier_sloan_2025} as part of the SPIDERS programme (SPectroscopic IDentification of EROSITA Sources; \citealt{dwelly_spiders_2017}; \citealt{comparat_final_2020}; \citealt{Aydar2025}) which aims to obtain optical/near-infrared spectra of sources detected by eROSITA. The SPIDERS targets in our sample were detected in either the full survey depth eFEDS region \citep[eROSITA Final Equatorial Depth Survey;][43 per cent of our sample]{brunner_erosita_2022, salvato_erosita_2022} or the first all-sky survey eRASS:1 \citep[32 per cent;][]{Merloni}. We supplement the eROSITA-selected sample with 695 sources from the \textit{Chandra} Source Catalog \citep{evans_2024} and another 55 from the combination of \textit{XMM-Newton} and \textit{Swift}-XRT serendipitous catalogues (\citealt{Webb_2020}; \citealt{Delaney_2023}),\footnote{\hli{The \textit{XMM-Newton} and \textit{Swift}-XRT catalogues were combined to form a single target catalogue for SDSS-V to assign spare fibres (i.e. as an ``open fiber program''), preferentially adopting the \textit{XMM-Newton} positions and X-ray fluxes, when available.}} all in the 1.5--3.5 redshift range and with S/N $\geq5$ per SDSS pixel. The \textit{Chandra} and \textit{XMM}/\textit{Swift} catalogues have been used to define additional targets for SDSS-V, probing a larger sky area and deeper X-ray observations. All spectra were reduced with the v6\_1\_3 version of the SDSS BOSS pipeline, idlspec2d (\citealt{bolton_2012}; Morrison et al., in preparation) as part of the SDSS DR19 \citep{collaboration_nineteenth_2025}. 

Our sample is X-ray selected by design and so will be affected by the sensitivity and depth of the X-ray instruments and surveys. The \textit{XMM} and \textit{Swift} detections (4.5--12 and 2--10\,keV, respectively) are at harder X-ray energies than eROSITA where most of the detections are below 2\,keV. Meanwhile the \textit{Chandra} sources were mostly detected at 0.5--7\,keV but observations are not uniform in depth. The largest subset of our sample is from eFEDS and so we consider the eFEDS flux limit ($10^{-14}$\,erg\,s$^{-1}$\,cm$^{−2}$) throughout the paper.

The redshift range 1.5--3.5 was chosen for examination because, at \textit{z} $<$ 1.5, the wavelength range of interest for {\CIV} BAL detection begins to shift out of the spectrograph's range, while above \textit{z} $>$ 3.5 the \ion{C}{iii}] emission line is absent which is necessary for spectral reconstructions (see section~\ref{sec:BAL identification}). The detectability of BAL-troughs decreases significantly as the S/N of a given spectrum decreases (\citetalias{gibson_catalog_2009}; \citealt{allen}). As we aim to investigate the BAL quasars relative to the non-BAL quasars in this X-ray-selected sample, a minimum spectrum S/N threshold is adopted to only include spectra with S/N $>$ 5.

\begin{figure*}
	\includegraphics[width=18cm,height=6cm]{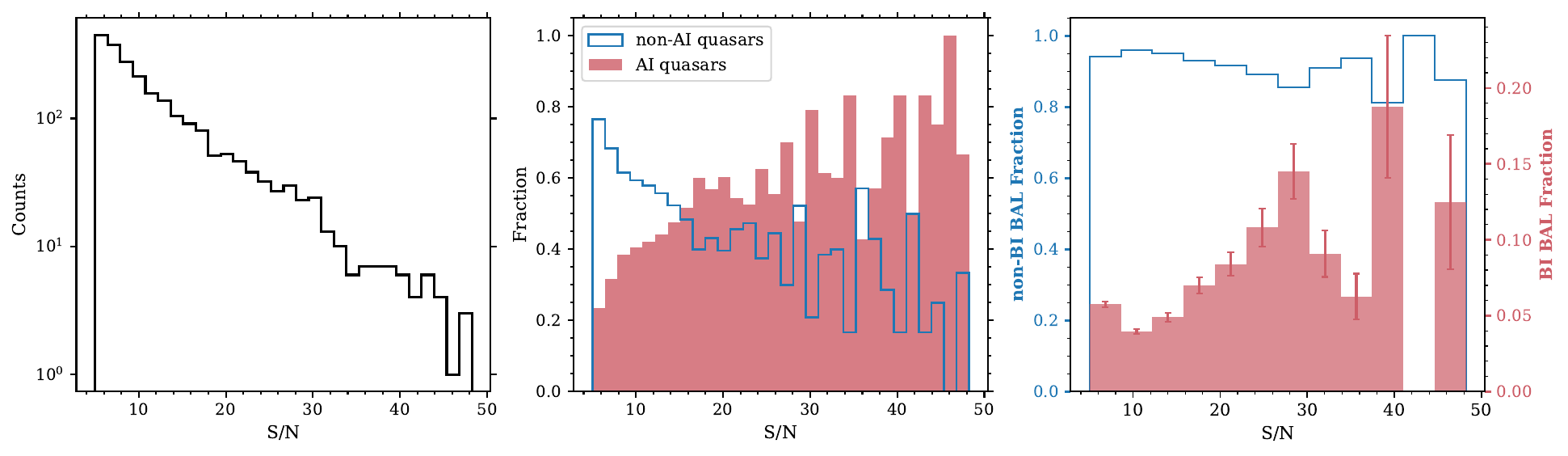}
    \caption{The \textit{first} panel shows the median S/N distribution in the SDSS spectra of our final X-ray selected sample numbering 2317 quasars, reduced from 2680 due to the removal of sources with suboptimal spectral reconstructions (see Section \ref{sec:BAL identification}). The \textit{second} and \textit{third} panels present the distribution of the fraction of AI quasars, BI BALs and non-BAL quasars as a function of median S/N. The fraction of BALs and non-BALs adds to unity in each bin (note the different axis for the BI BAL and non-BAL fraction in the \textit{third} panel). The fraction of BI BAL and AI quasars decreases at low S/N due to noisy spectra that make it harder to identify absorption features. 
    } 
    \label{fig:histograms}
\end{figure*}

The {\CIV} emission-line blueshifts that we are interested in cover velocities of order a few thousand {\kms} and so rely on accurate systemic redshifts. The BAL classifications are also dependent upon the exact redshift estimate for a quasar where the troughs lie close to the edges of the velocity window considered (see Section~\ref{sec:BAL identification}). We produced updated redshifts using the cross-correlation algorithm outlined in section 4.2 of \citet{hewett_improved_2010} and updated in \citet{stepney_no_2023}.  More information regarding the redshifts for the eROSITA-selected sub-sample can be found in Rankine et al. (in prep) and we adopt the same procedure for the additional \textit{Chandra}- and \textit{XMM/Swift}-selected sources. In practice, $>85$ per cent of the sample has redshift estimates that differ from the SDSS pipeline by $\vert\Delta z\vert<500$\,{\kms} and $<5$ per cent have $\vert\Delta z\vert>1000$\,{\kms}.

\hli{A table containing the optical and X-ray properties underpinning this paper is available as supplementary data.} Appendix~\ref{appendix_data} \hli{describes the data available.}

\section{BAL identification}\label{sec:BAL identification}

The first aim of this paper is to identify the BAL quasars within the X-ray-selected SDSS-V sample, and we have chosen to make this categorisation based on the {\CIV} emission line\hli{, while excluding quasars defined as LoBALs due to }\ion{Al}{III} and/or \ion{Mg}{ii}\hli{ absorption.} Two methods are commonly used to classify quasars as BAL or non-BAL: \citet{Weymann1991} defined the Balnicity Index (BI) while \citet{Hall_2002} introduced the Absorption Index (AI) to incorporate absorption close to systemic velocity and narrower troughs, often called mini-BALs. However, the AI metric can also contaminate BAL samples with strong narrow absorption lines (NALs) and blended NALs (see \citealt{knigge2008} for a full discussion and Appendix~\ref{appendix_spectra} for example spectra).
BI and AI are calculated as follows:
\begin{equation}\label{eq:BI}
     \text{BI} =  \int_{3000}^{25000} \left(1 - \frac{f(V)}{0.9}\right) C \,\text{d}V,
\end{equation} 
and
\begin{equation}\label{eq:AI}
\text{AI} =  \int_{0}^{25000} \left(1 - \frac{f(V)}{0.9}\right) C \,\text{d}V.
\end{equation}
$f(V)$ is the spectrum flux normalized by the continuum and $V$ is the velocity in km~s$^{-1}$ defined with respect to the quasar systemic velocity. In equation~\ref{eq:BI}, \hli{$C=1$ where $f(V)$ $< 0.9$ contiguously for at least 2000\,{\kms} and zero otherwise. The integration occurs between 3000 and 25\,000\,{\kms} from systemic velocity. Quasars with BI $>0$ are considered BAL quasars (BI $=$ 0 are non-BI BALs). In equation}~\ref{eq:AI}\hli{, $C=1$ where $f(V)$ $< 0.9$ contiguously for at least 450\,{\kms} and zero otherwise, integrated over 0--25\,000\,{\kms}. Quasars with AI $>0$ could either be mini-BALs or have strong NALs in their spectra; therefore, we do not refer to this population as BALs in this paper and instead denote them as ``AI quasars'' (AI $= 0$ are non-AI quasars). Quasars with BI and/or AI $= 0$ are collectively referred to as non-BALs in this paper.}
Given the narrower width requirement, we apply $\chi_{\textup{trough}}^2 \ge$ 10 in order to account for spurious troughs \citep{Trump2006} using:
\begin{equation}\label{eq:chisq_trough}
{\chi}_{\textup{trough}}^2 = \sum \frac{1}{N} \left(\frac{1 - f(V)}{\sigma} \right)^2,
\end{equation} 
where \textit{N} is the number of pixels in the potential false-positive AI troughs (due to noise) and $\sigma$ is the estimated noise for a given pixel. 

To obtain $f(V)$, we reconstruct the quasar spectra using the same procedure outlined in \citetalias{rankine_2020} using the same spectral components generated via mean-field Independent Component Analysis (ICA). The resulting reconstructions essentially fill in the BAL troughs to produce a smooth unabsorbed continuum, and this continuum is used to normalize the spectrum. The normalized spectrum, $f(V)$, is used for distinguishing it as a BAL or non-BAL using the BAL-identification code by \citet{Rogerson_2018}, where the code uses equations \ref{eq:BI}, \ref{eq:AI} and \ref{eq:chisq_trough}. 

We made use of the same ICA components as \citetalias{rankine_2020} that were generated using the SDSS-IV quasar sample of \citet{paris2018} \citepalias[see section 4.2 of ][]{rankine_2020}. The SDSS-V target selection differs greatly from that of previous SDSS surveys (for this paper, mainly due to targeting an X-ray selected quasar sample); thus it is not a foregone conclusion that the components would perform adequately as the SDSS-V quasar population may have significantly different optical/UV spectroscopic properties. We compare the quality of the reconstructions of the SDSS-IV and -V samples using the reduced $\chi^2$ of the fits. The use of the same components in reconstructing the SDSS-V spectra is justified due to the similarities in the distribution of the reduced ${\chi}^2$ values of the two samples as shown in Fig.~\ref{fig:chisqr}.

 
We limit our final sample to quasars that have spectral reconstructions with ${\chi}_r^2 < 2$ and place them in the {\CIV} emission line EW vs. blueshift space (see Section \ref{sec:CIVspace}) to visually inspect the objects lying in the extremes of {\CIV} emission space. Following \citetalias{rankine_2020}, we removed quasars with large negative {\CIV} blueshifts ($<$ -1200\,{\kms}), small {\CIV} EW ($<$ 10 \AA) and within the bottom left corner of the space (log$_{10}${\CIV}(EW) $<$ -2.3077$\times$ 10$^{-4}$ $\times$ {\CIV}(blueshift) + 1.321) due to suboptimal reconstructions of quasar spectra in this region. These considerations produced a final sample of 2317 quasars.

Figure~\ref{fig:histograms} shows that the fraction of BI BAL and AI quasars relative to non-BAL quasars in general increases with increasing S/N. This increase is steeper for the AI quasars due to the additional $\chi_{\text{trough}}^2$ requirement which introduces an additional bias against low-S/N spectra. Although the majority of the BI BAL quasars in this sample have low S/N, each spectrum has been visually inspected to confirm the presence of a broad absorption trough in the original spectrum. 
While there are relatively more BI BAL and AI quasars than non-BAL quasars in the high S/N bins, the first panel indicates that the overall number of quasar spectra in this high-S/N range is small. 

The observed fraction of BAL quasars in Fig.~\ref{fig:histograms} will be different from the intrinsic fraction due to selection effects resulting from the magnitude-limited nature of SDSS and S/N of the spectra. \citet{allen} quantified the effect of BAL quasars being redder (thus more likely to not satisfy the colour selection of previous SDSS surveys) and the effect of S/N (noise spikes halting the BI integration). SDSS-V target selection does not employ any colour constraints but does, for our sample, have an X-ray flux cut in its place. However, determining the intrinsic BAL fraction from this sample is beyond the scope of this paper (and is likely not feasible due to the complex way in which X-ray highly obscured systems \hli{are lost} from the sample); we instead aim to understand the BAL \textit{properties} of this new and large X-ray-selected BAL quasar sample. 

Previously studied optically selected samples have observed BAL quasars to be X-ray weaker relative to non-BAL quasars, so the initial aim of this paper is to identify BAL quasars within our SDSS-V sample to investigate their extent in the X-ray selected sample. Using the methods outlined in this section, we found our sample to consist of 143 or 6.2 per cent of BI BAL quasars \hli{(excluding 9 LoBALs)} and 954 or 41.2 per cent of AI quasars. 
\hli{Several studies find the observed BAL fraction to be in the range of 10\% to 15\% within optically selected quasar samples} (e.g., \citealt{Tolea_2002}, \citealt{Hewett_2003}, \citealt{knigge2008}, \citetalias{gibson_catalog_2009}).
\citet{Trump2006} \hli{quote BI and AI fractions of 10.4 and 60.6 per cent, respectively. Note that their definition of AI extends to 29\,000\,{\kms} (cf. our 25\,000\,\kms) and they also obtain an AI fraction of 26.0 per cent if they increase the minimum trough width from 450\,{\kms} to 1000\,{\kms}.} 
\hli{Thus, the fraction of BAL or AI quasars identified within our X-ray-selected quasar sample is comparable to---albeit slightly lower than---the fractions found in optically selected samples.}

In the following sections, we proceed to examine where the X-ray selected BAL quasars identified in our paper lie in the {\CIV} emission space compared to the non-BALs, and to explore their UV/optical and X-ray properties.

\section[C IV emission space]
{C\,{\sevensize IV} emission space}
\label{sec:CIVspace}

\begin{figure}
\centering
  \includegraphics[width=\linewidth]{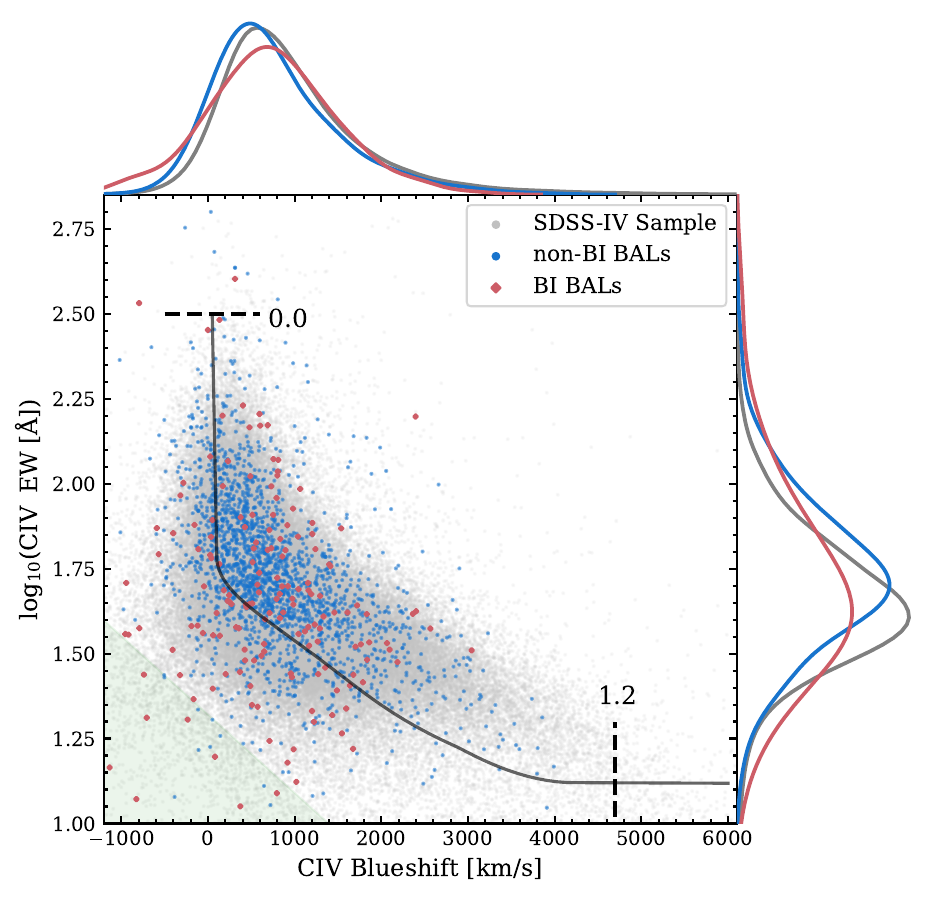}
  \includegraphics[width=\linewidth]{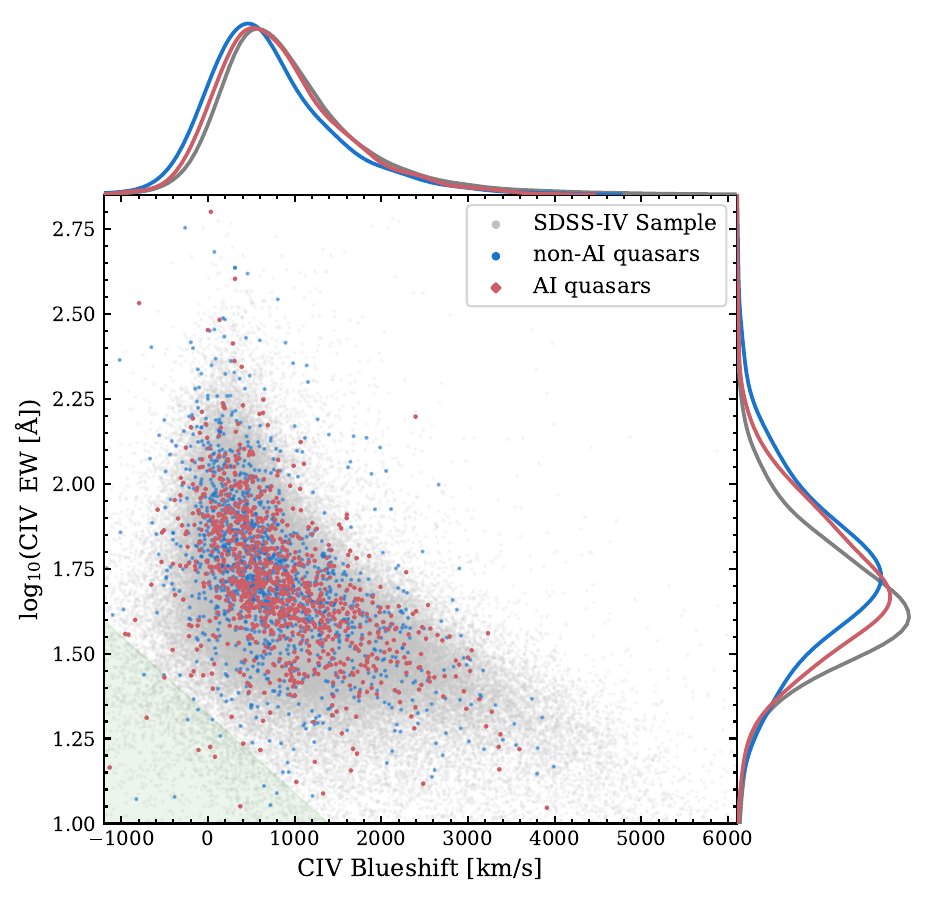}
\caption{The {\CIV} space with BI- (\textit{top panel}) and AI- (\textit{bottom panel}) defined quasars in red and non-BAL quasars in blue. These X-ray-selected BI BALs and AI quasars overlap with their non-BAL equivalents in the {\CIV} space. The optically selected SDSS-IV sample shown in grey in both panels (BALs and non-BALs) extends to higher {\CIV} blueshifts relative to the X-ray-selected sample in coloured dots. Additionally, the \textit{top panel} presents the best-fitting relation based on the SDSS-RM sample that is used to define {\CIV} distance that increases with decreasing EW and increasing blueshift. The green-shaded region in both panels represents the sources excluded from our sample due to suboptimal reconstruction of their spectra (see Section \ref{sec:BAL identification}).}
\label{fig:bal/nonbal}
\end{figure} 

As mentioned in Section~\ref{sec:intro}, \citet{Richards2011} demonstrated that the distribution of quasars within the {\CIV} emission space is useful for understanding which component -- disc (high {\CIV} EW and low blueshift) or wind (low {\CIV} EW and high blueshift) -- is dominant in the context of an accretion disc-wind model. \citetalias{rankine_2020} showed that optically selected BAL quasars exist with a range of {\CIV} emission properties (and were similar to the non-BAL quasars across various UV-derived properties). Hence, we employ this space to study the X-ray selected SDSS-V BALs relative to non-BAL quasars identified in this paper.

Both panels of Fig.~\ref{fig:bal/nonbal} show the optically selected SDSS-IV sample in grey dots \citepalias[from][and limited to the CORE SDSS-IV targets]{rankine_2020} relative to the SDSS-V sample in coloured dots, these samples are placed in the {\CIV} emission space where the EWs and blueshifts of the {\CIV} emission line have been measured non-parametrically using the ICA reconstructions, consistently to the methods used by \citetalias{rankine_2020}. The SDSS-IV sample is a significantly larger sample that extends to higher {\CIV} blueshifts relative to the X-ray-selected SDSS-V sample studied in this paper. While a more detailed comparison of the SDSS-IV and SDSS-V samples in the {\CIV} emission space will be presented by Rankine et al. (in prep), the difference in the extent of the {\CIV} blueshifts could be leading to the differences in their distributions, which can be seen via the 1D blueshift distributions in the same figure. Furthermore, a 1D Kolmogorov-Smirnov (KS) test indicates an extremely low probability that the SDSS-IV and SDSS-V samples are drawn from the same underlying distribution of EW ($p=3.2\times10^{-62}$) or blueshift ($p=1.2\times10^{-17}$) and thus we can conclude that they are significantly different. 

\citetalias{rankine_2020} reported that while the optically selected BAL quasars in their SDSS-IV sample are distributed across the entire {\CIV} space, they are most likely to have moderate to low EW and high blueshifts. In detail, the optically selected BAL quasars in the SDSS-IV sample are mildly skewed towards low EW and high blueshifts relative to the non-BAL quasars (see their figure 8). 
Figure~\ref{fig:bal/nonbal} demonstrates that the SDSS-V BI BAL and AI quasars are distributed across the {\CIV} space similarly to the non-BAL quasars. \hli{The small \textit{p}-values ($<10^{-5}$) resulting from KS tests, where we require $p <0.05$ to reject the null hypothesis of two populations being drawn from the same underlying distribution, suggest that the BAL and non-BAL populations are unlikely to be from the same underlying distribution.}

In an earlier study of BALs in the SDSS DR6 spectroscopic survey, \citet{allen} found an intrinsic redshift-evolving BAL fraction which results in an observed optical luminosity dependence when accounting for various selection effects. \citetalias{rankine_2020} described an optical luminosity dependence on {\CIV} location. \hli{Rankine et al. (in prep) will perform a more detailed analysis of the different redshift distributions; however, our sample does cover the same range of redshifts as} \citetalias{rankine_2020} and here we compile a redshift-matched BAL and non-BAL sample to check if the differences in the \textit{observed} redshift distribution between the BALs and non-BALs in our sample are leading to the differences in the {\CIV} emission space. For a given BI BAL quasar we chose three non-BI BAL quasars with the closest $z$ values, while for every AI quasar we chose one non-AI quasar with the closest $z$ value. We performed KS tests to compare the BAL  and the AI sample to their redshift-matched non-BAL equivalents in the {\CIV} space and report the resulting \textit{p}-values in Table \ref{tab:KStest}. 

\begin{table}
 \caption{KS test \textit{p}-values for the redshift matched BI and AI populations. The null hypothesis of the populations being drawn from the same distribution is rejected when $p<0.05$. X-ray properties of the BI/AI populations suggest that the BAL and non-BAL quasars are drawn from the same underlying distribution.}
 \label{tab:KStest}
 \begin{tabular}{lcc}
  \hline
  KS test & BI populations & AI populations\\
  \hline
  1D (EW) & 0.03 & 0.01 \\
  1D (Blueshift) & 0.5 & 0.03\\
  2D (EW-blueshift) & 0.02 & 0.01\\
  1D ($L_{\textup{X}}$) & 0.2 & 0.4\\
  2D ($L_{\textup{X}}$-z)& 0.3 & 0.4 \\
  1D ($N_\text{H}$) & 0.6 & 0.5\\
  1D ($\Delta \alpha_\text{ox}$) & \hli{0.4}  & \hli{0.8}\\
  \hline
 \end{tabular}
\end{table}

\begin{figure*}
	\includegraphics[width=18cm,height=7.5cm]{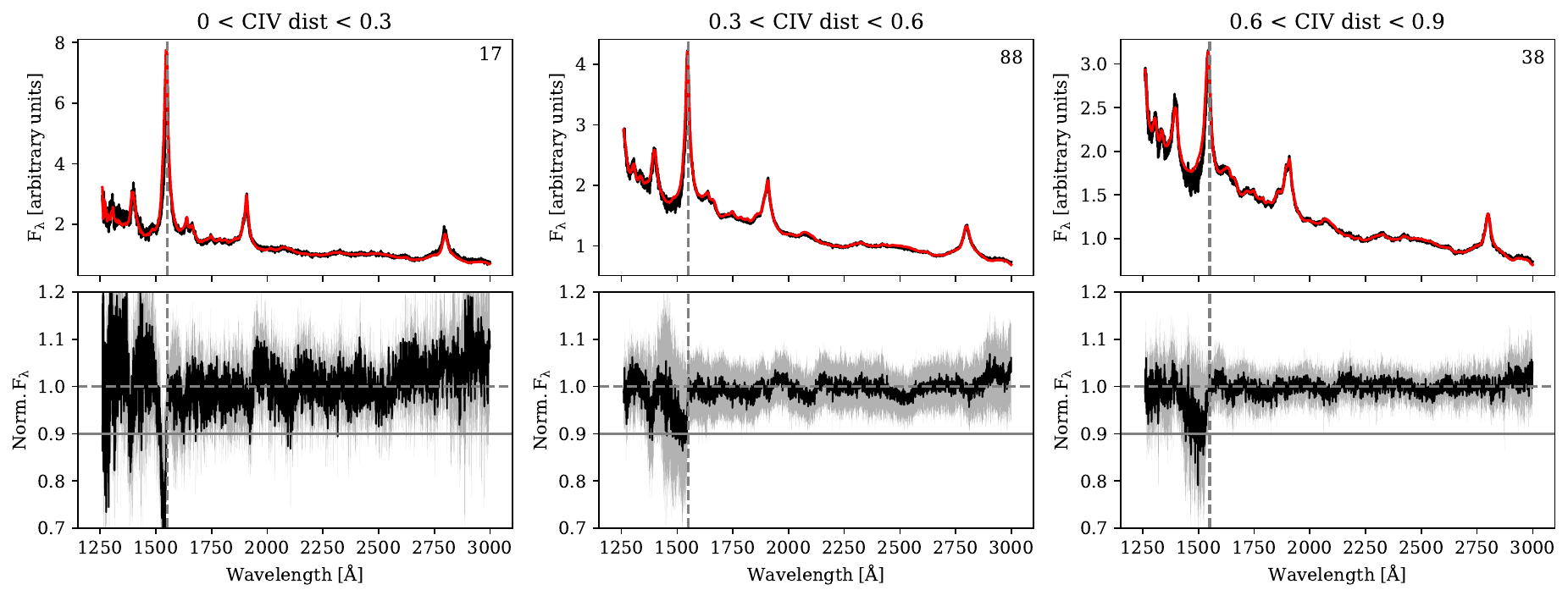}
    \caption{Median composite spectra of the BI BAL quasars in different {\CIV} distance bins with the \textit{top panels} showing the composite flux (black) and reconstruction (red) along with the number of sources that went into making the composites in the top right corner, the \textit{bottom panels} show composite spectra normalized by the reconstructions along with its Median Absolute Deviation (MAD) in grey. The solid horizontal grey line marks the 0.9 threshold for the BI calculation. Considering that our X-ray-selected BAL quasars do not extend to as high {\CIV} blueshifts as the SDSS-IV sample, expected absorption features are observed in our BI BAL quasars with the troughs becoming \hli{broader} with increasing {\CIV} distance.
    } 
    \label{fig:composites}
\end{figure*}

The blueshifts of the BI sample are statistically consistent with being drawn from the same distribution as the non-BI BALs ($p=0.5$), whereas the lower $p$-value when comparing the blueshifts of AI quasars to the equivalent non-AI quasars provide statistical significance ($p<0.05$) for rejecting the null hypothesis of being drawn from the same population; we therefore conclude that the underlying populations differ. 
We also find statistically significant $p$-values allowing rejection of the null hypothesis of the same underlying distributions when comparing the EWs of both the BI and AI samples to their non-BAL equivalents, as well as statistically significant $p$-values when performing a 2D KS test \footnote{The 2D KS tests were performed using the public code \textsc{ndtest} written by Zhaozhou Li, \url{https://github.com/syrte/ndtest}.} to compare the distributions of both BI BALs and AI quasars to their non-BAL equivalents in the two-dimensional EW--blueshift space. 
We conclude overall that the BAL samples have a distinct distribution in the \CIV\ EW--blueshift space compared to non-BALs that cannot be accounted for by differences in the observed redshift distributions. 
However, it is important to note that --- while statistically significant --- the differences are \emph{small} and appear mainly to be driven by the BALs being preferentially found in quasars with slightly lower \CIV\ EWs, 
i.e., BI BALs have median $\textup{log}_{10}$~(EW~[\r{A}]) $= 1.66\pm0.03$ compared to a median $\textup{log}_{10}$~(EW~[\r{A}]) $= 1.719\pm0.006$ for non-BI BALs; similarly, AI quasars have median $\textup{log}_{10}$~(EW~[\r{A}]) $=  1.697\pm0.008$ compared to $\textup{log}_{10}$~(EW~[\r{A}]) $= 1.738\pm0.008$ for non-AI quasars.
Nonetheless, BALs are distributed across the {\CIV} space occupied by the SDSS-V X-ray selected quasars.
We note that using the redshift-matched sample did not change the rest of our results so we retained the whole sample for our subsequent analysis for better number statistics.

Next, we investigate how the \emph{strength} of the BAL features (rather than simply their presence) varies across the {\CIV} emission space. 
We divide the BI BAL sample into different regions of the {\CIV} emission space based on their ``{\CIV} distance'', which is defined by \citet{richards_2021} as the distance along the best-fitting relation between {\CIV} EW and blueshift for the SDSS-RM sample from \citet{rivera_2020} and is shown by the black curve in the top panel of Fig.~\ref{fig:bal/nonbal}. In detail, each object in our sample is assigned a {\CIV} distance based on its closest position to this best-fitting relation to obtain its distance along the curve, using the code by \citet{CIVdist}. As noted in the figure, large {\CIV} distances imply low EWs and high blueshifts. Our sample typically lies to the right or above the \citet{richards_2021} curve due to slight differences in the way that the blueshift and EW are measured, but this does not affect the measurements of the {\CIV} distance.

\begin{figure}
	\includegraphics[width=\columnwidth]{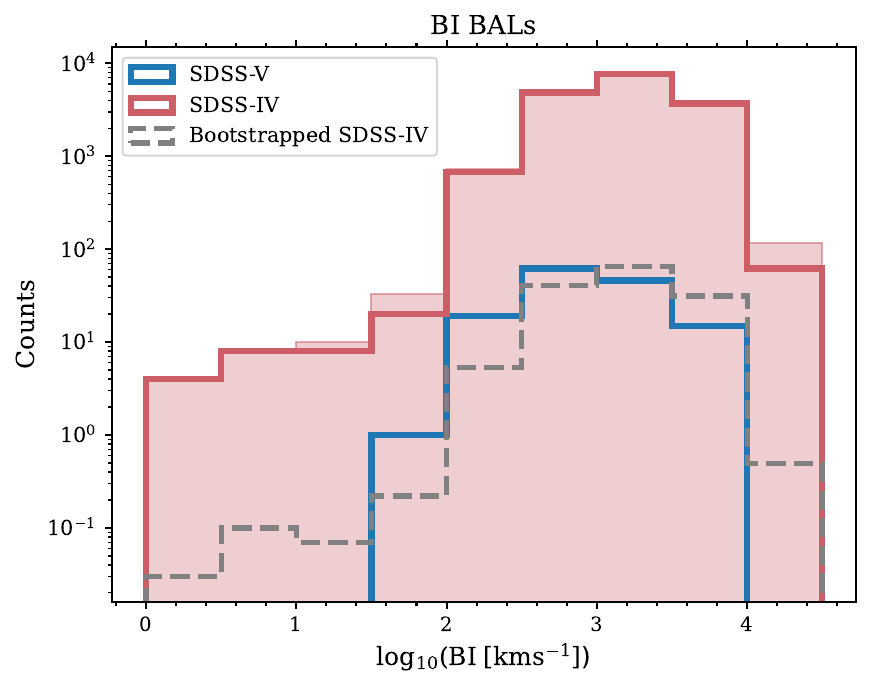}
    \caption{Comparing the BI values (in {\kms}, quantifying the strength of absorption in BALs, similar to an EW measure) of the BI-defined BAL quasars in the X-ray selected SDSS-V sample (in blue) and the optically selected SDSS-IV sample (in red). 
    The filled red histogram shows the full SDSS-IV sample, whereas the 
    open histograms are restricted to samples with -1000 {\kms} $\le$ {\CIV} blueshift $\le$ 3000 {\kms}, i.e., they are {\CIV} blueshift matched. 
    Our relatively smaller X-ray selected sample picks up the most common BI BALs, however, with a lack of objects with BI $>10^4$\,{\kms}. 
    \hli{The grey histogram shows a bootstrapped sub-sample from SDSS-IV, indicating that the differences in the shapes of the BI distributions of the SDSS-IV and -V BALs persists, most notably in the low and high BI tails.} The absence of BALs at these extremes in SDSS-V is likely due to its relatively lower S/N and X-ray selection.}
    \label{fig:bi_dist}
\end{figure}

Figure~\ref{fig:composites} displays the median composites of the X-ray-selected BI BAL quasars in different {\CIV} distance bins, with {\CIV} distance increasing from left to right. 
To generate the composites the original and the reconstructed spectra are shifted to the quasar rest frame and the median flux is calculated at each pixel for the original, reconstructed, and normalized (original/reconstruction) spectra. The composite spectra reveal that as the {\CIV} distance increases, the absorption features become
\hli{broader}. We remind the reader that the BI BAL sample does not extend to as high {\CIV} blueshifts/distances as the SDSS-IV sample from \citetalias{rankine_2020}, which results in \hli{no BI quasars in the highest {\CIV} distance bin ({\CIV} distance $>0.9$), and thus we do not observe the strongest BAL features, as per the correlation between emission and absorption signatures discussed in} \citetalias{rankine_2020}. Moreover, it was also observed that the absorption troughs for these BAL quasars were found to exist at a wide range of velocities (as observed in the median absolute deviation at each pixel --- the grey-shaded region in the bottom panels) such that the depth and width of many of the troughs are washed out in the composites. Figure~\ref{fig:bi_dist} reveals that our X-ray selected sample is indeed missing the quasars with the strongest absorption troughs (largest BI values) compared to the optically selected population. Given the known correlation between BI and the {\CIV} emission properties, we limit the optically selected sample to $-1000\leq$ {\CIV} blueshift $\leq3000$\,{\kms}. Removing the highest {\CIV} blueshifts objects is insufficient to account for the lack of objects with BI $>10^4$\,{\kms} in the X-ray selected sample. Our X-ray sample also lacks BI $<10^2$\,{\kms} objects. To test if the different distribution is purely due to our smaller sample size, we perform bootstrap sampling by randomly sampling 143 BI BALs from the SDSS-IV sample 100 times to find the average distribution shown as the dashed grey histogram in Fig.~\ref{fig:bi_dist}. \hli{While the SDSS-V and the bootstrapped SDSS-IV BI distributions are similar between $1.5 < \textup{log}_{10}$ (BI [\kms]) $ < 4$, the differences between the distributions persist. Our sample is also missing the lowest BI objects, likely due to the lower average S/N of our spectra compared to the SDSS-IV sample.}

\begin{figure}
    \includegraphics[width=\columnwidth]{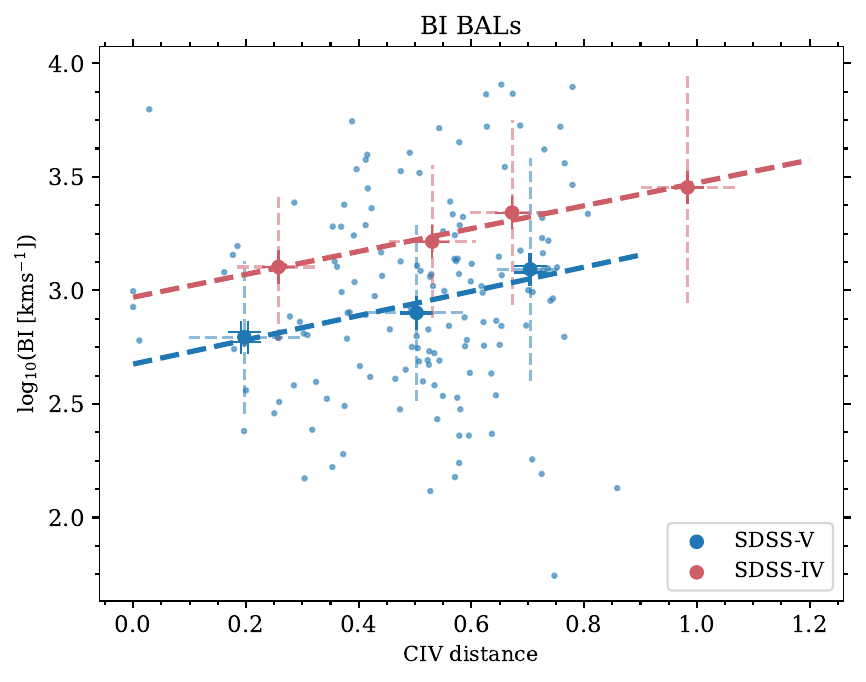}
    \includegraphics[width=\columnwidth]{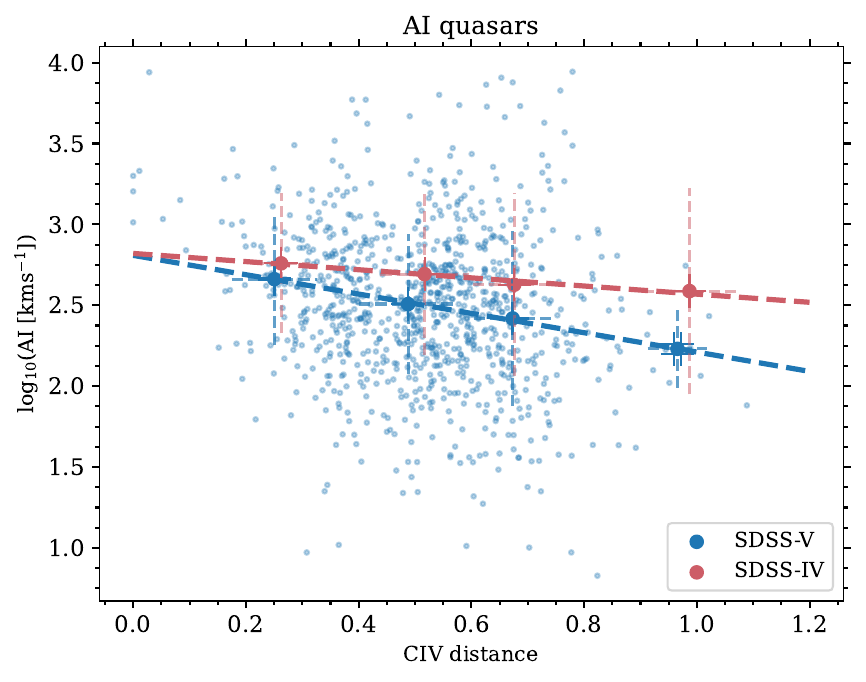}
    \caption{The relations of BI (\textit{top panel}) and AI (\textit{bottom panel}) against {\CIV} distance with the blue scatter points displaying all of the BI- and AI-defined quasars. The blue dots show the median BI and AI values in {\CIV} distance bins for the quasars identified in the SDSS-V sample (the same bins used in Fig.~\ref{fig:composites} for BI and Fig.~\ref{fig:composites_ai} for AI). Similarly, the red dots indicate the BI- and AI-defined quasars in the optically selected SDSS-IV sample. The dashed lines show the best fit for the medians, the dashed error bars indicate the standard deviation, and the capped error bars are the standard errors in the medians.
    The median behaviour in the X-ray selected BI BAL (AI) quasars in our sample is such that the absorption increases (decreases) with increasing {\CIV} distance.
    The SDSS-IV sample has systematically higher BI and AI values relative to the SDSS-V sample, although both samples show a high level of overlap due to the large scatter.} 
    \label{fig:bi_vs_civ}
\end{figure}

Figure~\ref{fig:bi_vs_civ} uses the same {\CIV} distance bins used in Fig.~\ref{fig:composites}, to examine how the median BI values (from Equation \ref{eq:BI}) change as the {\CIV} distance increases, where the BI value provides a measure similar to an EW measure (in {\kms}) of the broad absorption features that are $\geq$ 2000 {\kms} wide blueward of the {\CIV} emission line. These median BI values, shown as the blue dots in the figure, suggest an increase in absorption as the {\CIV} distance increases for the X-ray-selected BI BAL quasars in our sample, as previously noted via the composite spectra. While this trend agrees with the trend observed in the optically selected SDSS-IV sample studied by \citetalias{rankine_2020} (median values shown as the red dots), we note a substantial scatter, i.e., the dashed error bars show the standard deviation of the BI values in each {\CIV} distance bin which suggests the presence of a broad scatter. However, the median BI values are well constrained as indicated by the standard errors in the medians shown using the solid error bars, i.e., the median values increase from 2.79 $\pm$ 0.02 at 0.2 {\CIV} distance to 3.01 $\pm$ 0.01 at 0.7 {\CIV} distance, indicating a significant increase in the average strength of absorption within BI BAL quasars associated with strong {\CIV} outflows. While the SDSS-IV sample has systematically higher BI values relative to the SDSS-V sample, they are consistent within the errors.

\begin{figure*}
	\includegraphics[width=18cm,height=7.5cm]{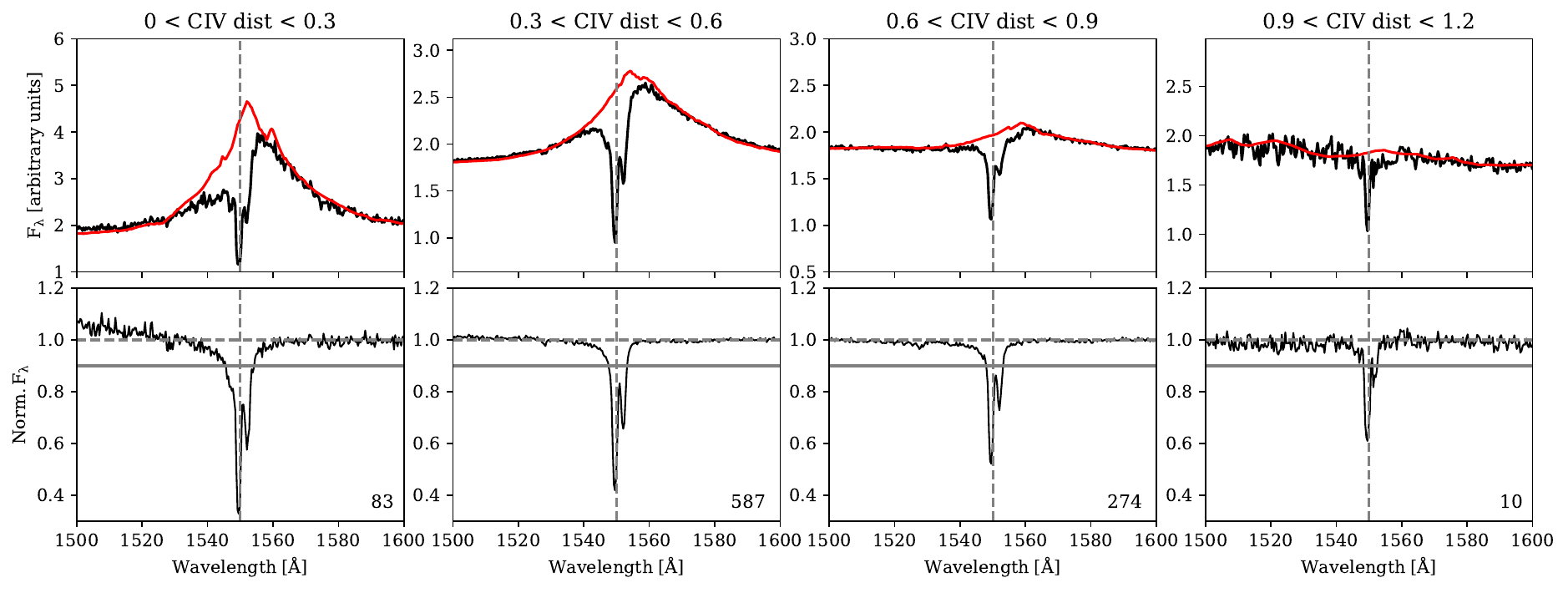}
    \caption{Same as Fig.~\ref{fig:composites} but for the AI quasars and in the rest frame of the absorber (the MAD is also omitted for clarity). The X-ray-selected AI quasars show the presence of NALs that decrease in strength at higher {\CIV} distance (see also bottom panel of Fig.~\ref{fig:bi_vs_civ}).}
    \label{fig:composites_ai}
\end{figure*}

The AI metric, however, is seen to decrease with increasing {\CIV} distance (bottom panel of Fig.~\ref{fig:bi_vs_civ}), in contrast to the increasing BI values with this distance.
The AI definition allows the inclusion of mini-BALs \citep[velocity widths of $\sim$1000--2000\,{\kms};][]{Trump2006} as well as strong narrow absorption lines (NALs: FWHM $<500$\,\kms; e.g., \citealt{hamann_elemental_1999, Bowler2014} -- see Appendix \ref{appendix_spectra} for representative rest frame spectra). Composite spectra of the AI quasars in the quasar rest frame (similar to Fig.~\ref{fig:composites}) contained no evidence of absorption due to the narrow width and broad velocity range of the AI-defined troughs. We instead generate composite spectra in the absorber rest frame, $z_\text{abs}$ as defined in equation 1 of \citet{Hall_2002} (introduced in \citet{weymann_results_1979} \& \citet{foltz_c_1986}):
\begin{equation}\label{eq:velocity}
    \beta = \frac{v}{c} = \frac{R^2 - 1}{R^2 + 1}; R = \frac{1 + z}{1 + z_\text{{abs}}},
\end{equation}
where $z$ is the redshift of the quasar, $c$ is the speed of light, and $v$ is the velocity of the deepest absorption relative to the {\CIV} line. The composites are presented in Fig.~\ref{fig:composites_ai} and show a clear {\CIV} NAL doublet. The absorbers producing these features have very different internal kinematics compared to absorbers that produce BI-defined absorption troughs. In addition to outflowing absorbers, some fraction are likely to be a consequence of intervening material along the line-of-sight to the quasar. The decreasing strength of the absorption with increasing {\CIV} distance is also evident in the normalized spectra (bottom panels). This differing behaviour of the BI- and AI-defined troughs in {\CIV} space was also observed for the optically selected sample in \citetalias{rankine_2020}.

\begin{figure}
	\includegraphics[width=\columnwidth]{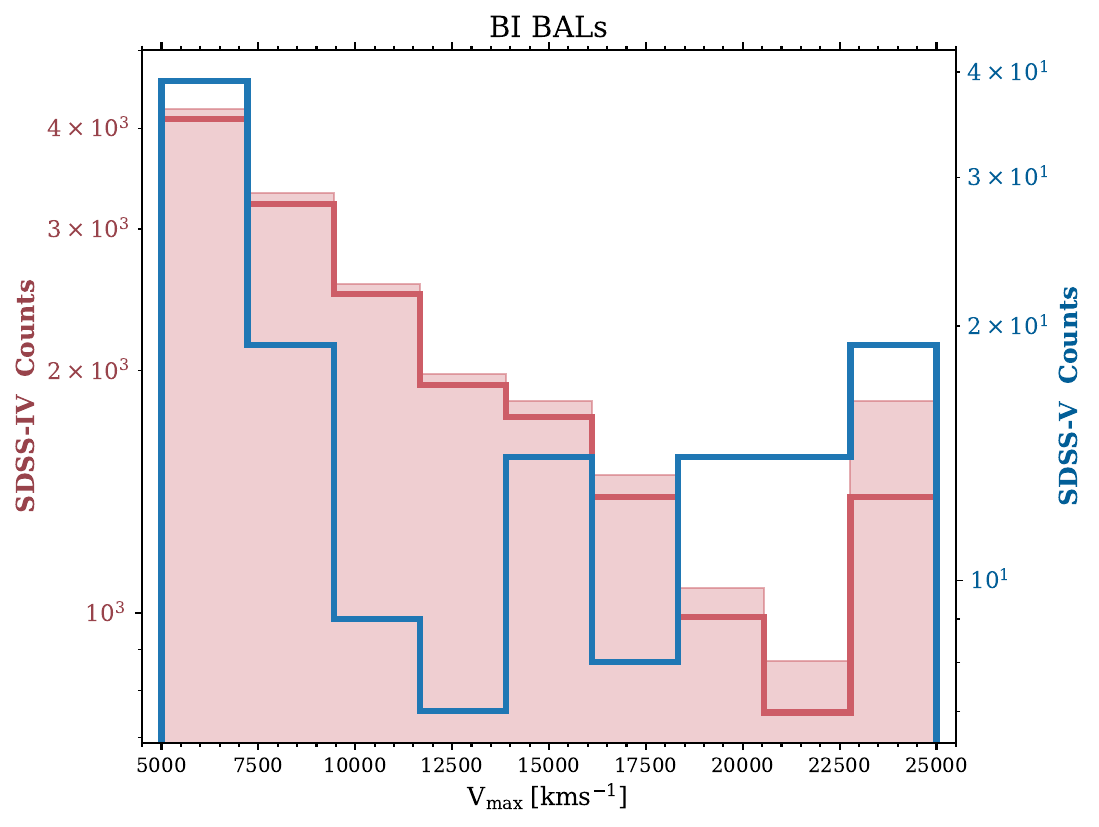}
    \caption{Comparing the maximum velocity at which the {\CIV} absorption or the $\textup{V}_{\textup{max}}$ occurs for the BI BAL quasars in the X-ray selected SDSS-V sample (in blue)and the optically selected SDSS-IV sample (in red). 
    The red histogram shows the full SDSS-IV sample.
    The step-style histograms compare $\textup{V}_{\textup{max}}$ of the {\CIV} blueshift matched SDSS-IV and SDSS-V samples, i.e., BI BAL quasars between -1000 {\kms} $\le$ {\CIV} blueshift $\le$ 3000 {\kms}.
    These BI BALs show that regardless of optical/X-ray selection BAL absorptions occur at a full range of velocities, in contrast to the BAL {\CIV} emission line velocities or blueshifts that show dependence on the wavelength selection.} 
    \label{fig:vmax_dist}
\end{figure}

\begin{figure}
	\includegraphics[width=\columnwidth]{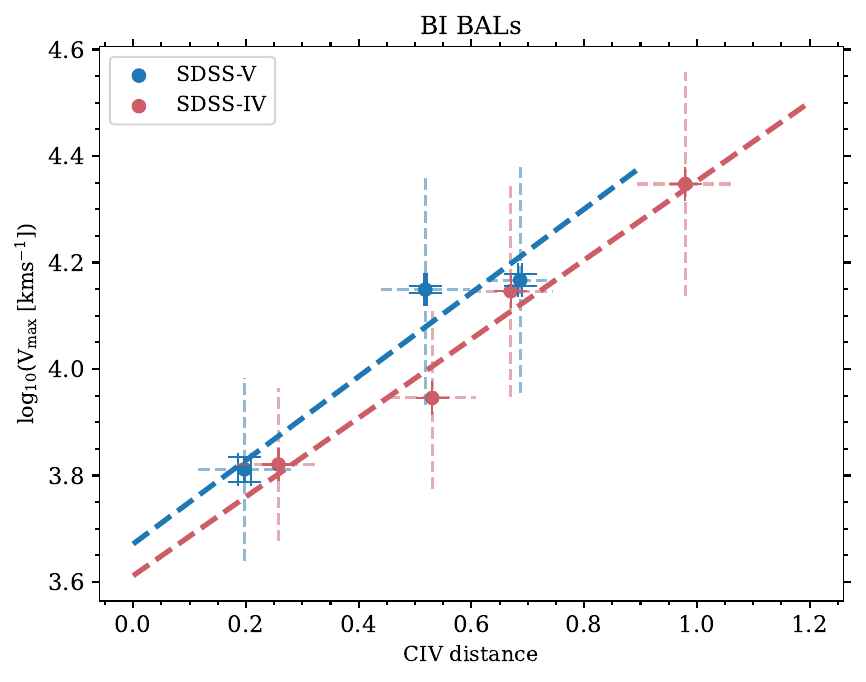}
    \caption{The median $\textup{V}_{\textup{max}}$ is observed to increase with increasing {\CIV} distance for both the BI BAL SDSS-V quasars (in blue) and SDSS-IV (in red). Where the error bars are the same as in Fig.~\ref{fig:bi_vs_civ} and the median $\textup{V}_{\textup{max}}$ is determined in {\CIV} distance bins of size $\Delta${\CIV} distance = 0.3 between 0 $\le$ {\CIV} distance $\le$ 1.2. The dashed lines show the best fit for the medians. This increase in $\textup{V}_{\textup{max}}$ with {\CIV} distance suggests that while the extent of the {\CIV} blueshifts, i.e., the emission velocities of the optically and X-ray samples differ (the BI BALs in the SDSS-V sample do not have sources in the last {\CIV} distance bin, see Fig.~\ref{fig:bal/nonbal}), the absorption velocities are relatively consistent at a given {\CIV} distance.} 
    \label{fig:vmax}
\end{figure}

Figure~\ref{fig:vmax_dist} shows the distribution of the maximum velocity at which the broad absorption blueward of the {\CIV} emission line, referred to as $\textup{V}_{\textup{max}}$ (in {\kms}), occurs in the BI BAL SDSS-IV and SDSS-V quasars. The open histograms compare the {\CIV} blueshift-matched optically selected and X-ray selected BI BAL quasars revealing that the absorption velocities span the full range independent of the selection. The pile-up at 25000\,\kms in this figure is explained by considering the BI definition in Equation \ref{eq:BI}, where the upper limit on the integral causes the troughs that extend beyond this limit to be truncated. Moreover, the full SDSS-IV sample in red (shaded) matches the range of $\textup{V}_{\textup{max}}$ of the SDSS-V sample, indicating that while the {\CIV} emission velocities of the samples show dependence on selection (Fig.~\ref{fig:bal/nonbal}), the range of absorption velocities does not. This is further explored in Fig.~\ref{fig:vmax} that shows the dependence of $\textup{V}_{\textup{max}}$ on {\CIV} distance, with both samples sharing a common trend of increasing $\textup{V}_{\textup{max}}$ with distance, consistent with the trend in composite spectra as previously shown in Fig.~\ref{fig:composites}. Therefore, this implies that while the extent of the {\CIV} blueshifts differs in the optical and X-ray selected samples, i.e., the emission velocities differ, the absorption velocities appear to match with each other. The similar range of $V_\text{max}$ combined with the changing median $V_\text{max}$ with {\CIV} distance suggest that the \textit{shape} of the $V_\text{max}$ distribution is changing with location in {\CIV} space. This is a new result and applies to both the optically and X-ray-selected populations. 

\section{X-ray properties of BAL quasars}\label{sec:X-ray prop}\begin{table}
\caption{Origin of the X-ray data in our sample.} 
\label{tab:xray}
\begin{tabular}{ccc}
    \hline
    Instrument & Observed energy band & Number of sources\\
               & keV & \\
    \hline
    eROSITA (eFEDS) & 0.2--2.3 & 959\\
    eROSITA (eRASS:1) & 0.2--2.3 & 716\\
    \textit{XMM} & 4.5--12 & 10\\
    \textit{SWIFT} &  2--10 & 12\\
    \textit{Chandra} & 0.5--7 (ACIS broad) & 508\\
                     & 1.2--2 (medium) & 14\\
                     & 0.5--1.2 (soft) & 1\\
                     & 0.1--10 (HRC broad) & 2\\
    \hline
\end{tabular}
\end{table}

\begin{figure}
\centering
  \includegraphics[width=\linewidth]{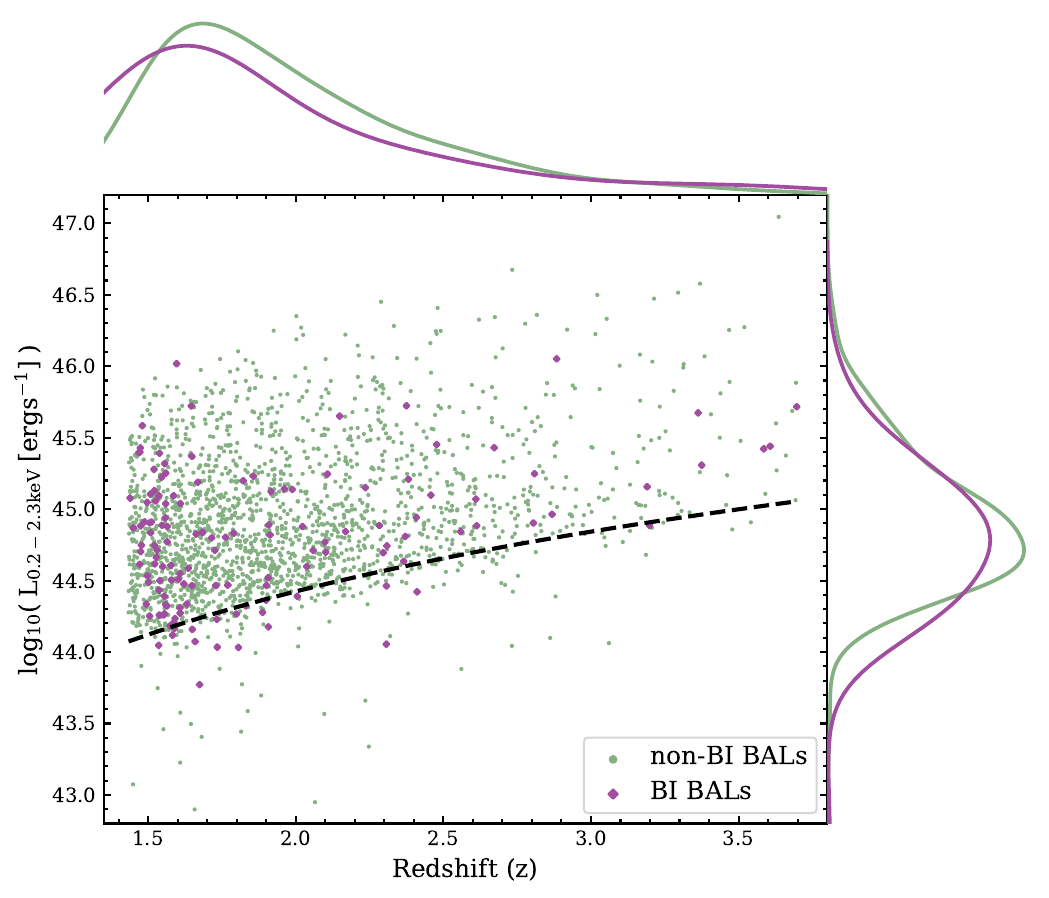}  \includegraphics[width=\linewidth]{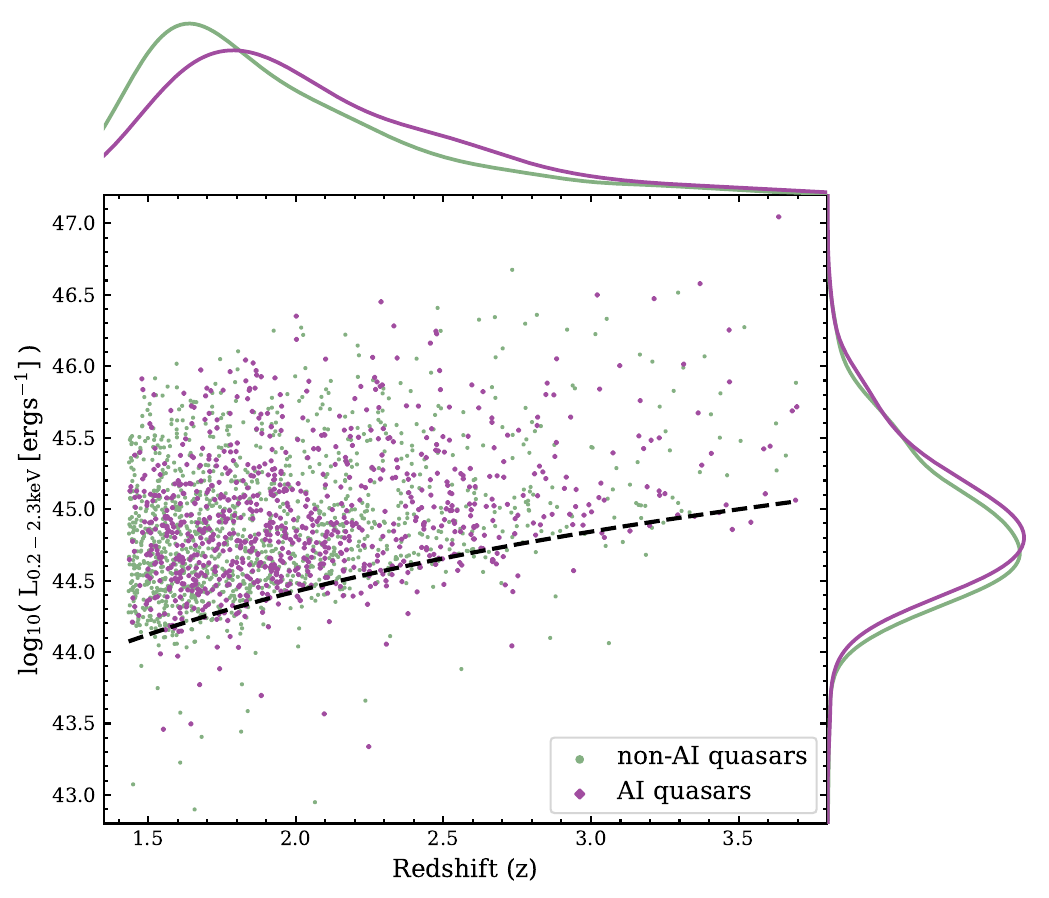}
\caption{The observed X-ray luminosities of the BI and AI populations. The purple diamonds represent the BI BAL and AI quasars and the green dots represent the non-BAL quasars, along with the black dashed line showing the luminosity at the limiting flux (\hli{adopting the eFEDS flux limit of $10^{-14}$\,erg\,s$^{-1}$\,cm$^{-2}$}; \citealt{brunner_erosita_2022}). Note that the size of BI BAL points has been increased for visibility. The distribution of the X-ray luminosities of all the sub-populations is similar over redshift, consistent with the previous result of finding these populations to be overlapping in the {\CIV} space (see Fig.~\ref{fig:bal/nonbal}). The X-ray selected BI BAL quasars do not appear X-ray weaker relative to the non-BALs, i.e., by construction, our sample is probing the X-ray bright BAL quasars.}
\label{fig:xlum}
\end{figure} 

In this section, we study the X-ray properties of the BAL and non-BAL quasars in our X-ray-selected sample to investigate the X-ray weakness of BAL quasars. 
The rest-frame 0.2--2.3\,keV X-ray luminosities  ($L_\textup{X}$) of our sample were calculated assuming a photon index of $\Gamma = $ 1.9 and using fluxes from eFEDS (43\% of the sample), eRASS:1 (32\%), \textit{XMM/Swift} ($\sim 1\%$) or \textit{Chandra} (23\%) in descending order of preference for objects with matches in multiple X-ray catalogues.
We adopted the fluxes measured in the 0.2--2.3\,keV band for the eROSITA (eFEDS and eRASS:1) sources, the 4.5--12\,keV band for the sources identified by \textit{XMM}, the 2--10\,keV band for the sources from \textit{Swift}, while for \textit{Chandra} sources we use fluxes measured in the 
ACIS broad band (0.5--7\,keV), medium band (1.2--2\,keV), soft band (0.5--1.2\,keV), and HRC broad band (0.1--10\,keV) 
as provided in the \textit{Chandra} Source Catalog \citep{evans_2024}. See Table~\ref{tab:xray} for a summary. The catalogue fluxes were calculated assuming Galactic absorption values of $N_\text{H} = 3 \times 10^{20}$\,cm$^{-2}$ for eROSITA sources \citep{brunner_erosita_2022, Merloni} and \textit{XMM} \citep{rosen_xmm-newton_2016}, and $4\times10^{20}$\,cm$^{-2}$ for \textit{Swift} sources \citep{Delaney_2023}. 
Galactic $N_\text{H}$ was estimated for each \textit{Chandra} source individually \citep{evans_2024}.

Figure~\ref{fig:xlum} shows that the $L_\textup{X}$ distribution of the BI- and AI-defined quasars over redshift is similar to the non-BAL quasars, suggesting that the X-ray selected BI BAL and AI quasars do not appear to be X-ray weaker relative to the non-BAL quasars in the sample studied in this paper (cf. \citealt{green_1995}; \citealt{laor_1997}; \citealt{brandt_2000}; \citealt{gallagher_exploratory_2006}; \citetalias{gibson_catalog_2009}; \citealt{Luo2014}; \citealt{saccheo2023}). 
This similarity in the $L_\textup{X}$ values may be due to the fact that we are working with an X-ray-selected sample and thus our selection is biased towards the X-ray brightest BAL quasars. Thus, our sample -- by construction -- will only probe the X-ray bright tail of the underlying BAL quasar population. \hli{The majority of the objects below the eFEDS flux limit are from the deeper \textit{Chandra} sample.\footnote{We adopt the eFEDS flux limit as the bulk of our X-ray selected sample were identified with eROSITA (from eRASS:1 or eFEDS catalogues) and the eFEDS flux limit corresponds to the deeper portion of this imaging.}}
The similar X-ray luminosities of the X-ray selected BAL and non-BAL quasars are consistent with our finding that the distribution of these two populations in the {\CIV} space are also similar (see Fig.~\ref{fig:bal/nonbal} and Section \ref{sec:CIVspace}), 
suggesting that -- for an X-ray-selected sample -- the BAL and non-BAL quasars are indistinguishable in their characteristics except for the broad (or narrow) absorption troughs that identify the BI BAL and AI quasars.
Additionally, the KS test \textit{p}-values presented in Table \ref{tab:KStest} for $L_\textup{X}$ and $L_\textup{X} - z$ space of the BI/AI populations, also imply that the BALs and non-BALs in our sample are drawn from the same underlying distribution.
\begin{figure}
\centering
  \includegraphics[width=\linewidth]{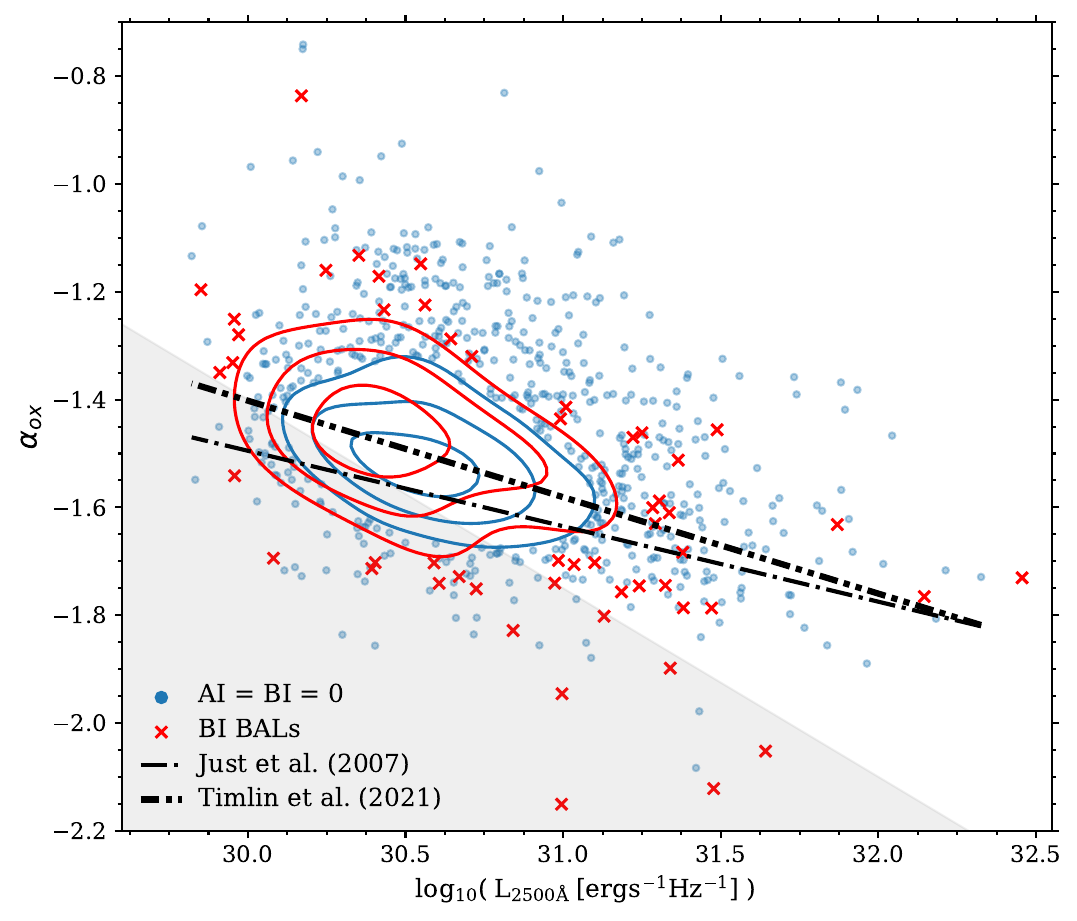}  \includegraphics[width=\linewidth]{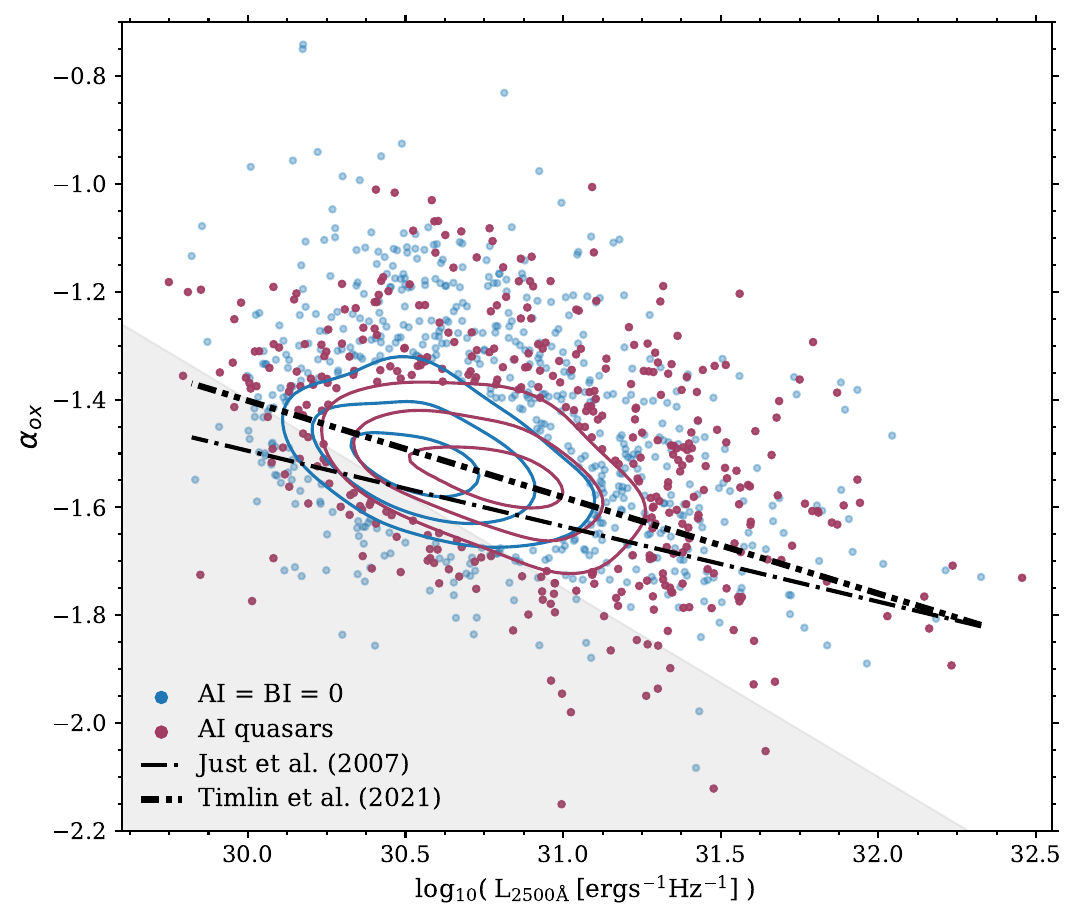}
\caption{\hli{$\alpha_\text{ox}$ vs $L_{2500}$ for our sample of X-ray selected SDSS-V quasars with the non-BALs (AI = BI = 0) in blue, the BI BALs in red (\textit{top panel}) and the AI quasars in dark red (\textit{bottom panel}). The contours are defined at 10\%, 30\% and 50\% of each of the samples. The BI BAL sub-population overlaps with the AI = BI = 0, suggesting that the X-ray weakness of a quasar is independent of its categorization as a BAL or non-BAL. Similarly, the AI quasars overlap with the AI = BI = 0 defined non-BAL population.}
The grey shaded region shows the eFEDS flux limit, with the \citet{just_2007} and \citet{Timlin2021} relations as the black dotted/dashed lines. The relations have slopes shallower than the sample studied in this paper, with the flux limit of our sample introducing a tilt in its distribution and missing the X-ray weaker (i.e. lower $\alpha_\text{ox}$) population at lower optical luminosities.}
\label{fig:alpha_ox}
\end{figure} 

This result motivates us to explore in further detail the observed X-ray properties of the BAL quasars relative to non-BALs. Hence, we employ $\alpha_\text{ox}$ to investigate the observed lack of difference in $L_\textup{X}$ values of these two sub-populations.
As discussed in Section \ref{sec:intro}, the UV/optical and X-ray luminosities of AGN are related (e.g., \citealt{Avni1982}), and this relationship is often quantified using the spectral slope, $\alpha_\text{ox}$, between the optical and X-ray as
\begin{equation}\label{eq:alpha_ox}
   \alpha_\text{ox}=\frac{\log_{10}\left(L_\textup{2keV}/L_{\textup{2500 \AA}}\right)}{\log_{10}\left(\nu_{\textup{2keV}}/\nu_{\textup{2500 \AA}}\right)},
\end{equation} 
where $L_{2500\text{\AA}}$ is the monochromatic optical luminosity at 2500\,\AA\ and $\nu_{2500\text{\AA}}$ is the frequency corresponding to this wavelength, while $L_{2\text{keV}}$ and $\nu_{2\text{keV}}$ are, equivalently, the monochromatic X-ray luminosity at 2\,keV and corresponding frequency. The monochromatic X-ray luminosities were calculated by converting full-band luminosities assuming a photon index of $\Gamma$ = 1.9 \citep[e.g.,][]{nandra_xmm-newton_2007}. Figure~\ref{fig:alpha_ox} shows the calculated $\alpha_\text{ox}$ as a function of $L_{2500\text{\AA}}$ for the BI- and AI-defined sub-populations in our X-ray-selected sample. All quasars in our sample show the expected anti-correlation between $L_{2500\text{\AA}}$ and $\alpha_\text{ox}$, which suggests that while optically brighter quasars are more X-ray luminous, overall, the \emph{relative} strength of the X-ray emission (compared to the optical emission) decreases at higher optical luminosities. While the contours for the BI BAL and AI quasars show that these quasars defined in our X-ray selected sample overlap with each other, they also overlap with the non-BAL quasars consistent with our previous result finding these sub-populations having similarly distributed $L_\text{X}$.
 
This result further suggests that the X-ray selected BI BAL quasars show a lack of X-ray weakness relative to the non-BAL quasars with all of the sub-populations showing similar UV/optical and X-ray properties. This conclusion is further motivated by the \textit{p}-values presented in Table \ref{tab:KStest} for a KS test comparing the $\alpha_\text{ox}$ distributions.

$\alpha_\text{ox}$ is a function of $L_{2500\text{\AA}}$ for quasars and \citet{just_2007} found this correlation to be
\begin{equation}\label{eq:alpha_ox_fit}         
    \alpha_\text{ox}(L_{2500\text{\AA}})= (-0.140\pm0.007)\;\textup{log}(L_{2500 \text{\AA}}) + (2.705\pm0.212),
\end{equation}
for a sample of optically bright SDSS \citep{York_2000} quasars within the redshift range of 1.5 $\le z \le$ 4.5 with additional complementary high luminosity quasars with $z \ge$ 4, and this correlation is shown as the black dashed line in Fig.~\ref{fig:alpha_ox}. This figure also shows a more recent linear anti-correlation equation found by \citet{Timlin2021} that made use of the same sample as \citet{just_2007} with additional restrictions of requiring the quasars to be within 1.6 $\le z \le$ 3.5, be radio-quiet, and devoid of BAL features: 
\begin{equation}
    \alpha_\text{ox}(L_{2500\text{\AA}}) = -0.179\; \textup{log}(L_{2500\text{\AA}}) + 3.968.
\end{equation} 

\begin{figure}
	\includegraphics[width=\columnwidth]{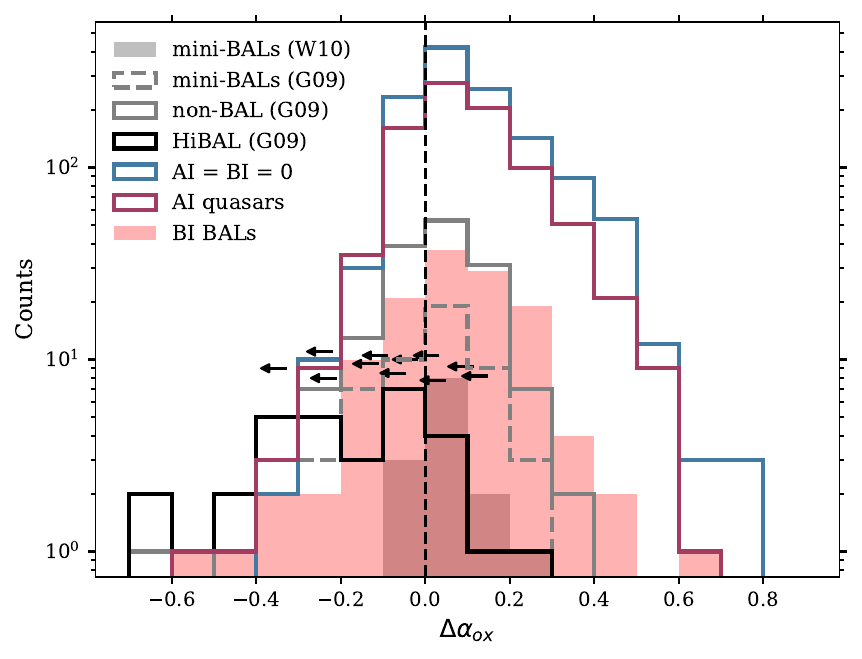}
    \caption{The distribution of $\Delta \alpha_\text{ox}$ for the BI BALs, AI quasars and non-BAL quasars (following the same colour scheme as in Fig.~\ref{fig:alpha_ox}), with the vertical dashed line indicating where $\Delta \alpha_\text{ox}$ = 0 occurs. 
    The grey histogram shows the optically bright mini-BALs studied by \citetalias{wu_x-ray_2010}.
    The optically selected HiBALs, mini-BALs and non-BAL samples studied by \citetalias{gibson_catalog_2009} are shown as the black, dashed grey and grey step-style histograms, respectively.
    \hli{The arrows indicate the 1$\sigma$ upper limits of the non-detections in the }\citetalias{gibson_catalog_2009} \hli{sample (the y-coordinates of these arrows are arbitrary).}
    The relatively high $\Delta \alpha_\text{ox}$ values of our sample are expected since X-ray selection biases us towards X-ray brighter sources at the optical luminosities probed by SDSS-V. 
    The BI BAL quasars peak around $\Delta \alpha_\text{ox}=0.2$ and have a broad distribution extending between $- 0.5$ $\lesssim \Delta \alpha_\text{ox} \lesssim$ 0.6. 
    These findings indicate that we probe the high $\Delta \alpha_\text{ox}$ region more fully and identify a relatively X-ray bright BAL population.}
    \label{fig:delta_ox}
\end{figure}

Figure~\ref{fig:alpha_ox} demonstrates that the \citet{just_2007} and \citet{Timlin2021} slopes are much shallower relative to the sample studied in this paper due to their differing sample selection. In this paper, however, we focus on comparing the properties of BALs and non-BALs in our new X-ray-selected sample. The grey-shaded region in Fig.~\ref{fig:alpha_ox} indicates the impact of the eFEDS flux limit in this space, evaluated at the redshift of our full quasar sample (1.5 $\le z \le$ 3.5) and thus how the X-ray selection will miss the X-ray weaker (i.e., lower $\alpha_\text{ox}$) quasar population at lower optical luminosities. 
\hli{Again, the majority of the sources below the flux limit are from the \textit{Chandra} sample.} The flux limit also introduces a tilt in the distribution of our samples in this space that leads to a deviation in the \emph{apparent} slope of the $\alpha_\text{ox}$-$L_{2500\text{\AA}}$ sample compared to the underlying relations measured by \citet{just_2007} and \citet{Timlin2021}.
By measuring the offset from these correlations,
\begin{equation}\label{eq:daox}
    \Delta \alpha_\text{ox} = \alpha_\text{ox} - \alpha_\text{ox}(L_{2500\text{\AA}})
\end{equation}
where $\alpha_\text{ox}(L_{2500})$ is the predicted $\alpha_\text{ox}$ at a given $L_{2500}$ based on the \citet{just_2007} best-fitting relation, we can compare the relative X-ray brightness across the full optical luminosity range between the X-ray selected BAL quasars and the non-BAL quasars.
A lower $\Delta \alpha_\text{ox}$ value implies X-ray weakness, i.e., if $\Delta \alpha_\text{ox} = −0.4$, then the quasar is X-ray weaker by a factor of $\approx$ 11 with respect to quasars with similar UV luminosities \citepalias{gibson_catalog_2009}. \hli{We define X-ray strong as $\Delta \alpha_\text{ox}>0.2$ and X-ray weak as $\Delta \alpha_\text{ox}<-0.2$.} Figure~\ref{fig:delta_ox} shows the distribution of $\Delta \alpha_\text{ox}$ for all the defined sub-populations relative to the \citet{just_2007} relation. The relative skew of the $\Delta\alpha_\text{ox}$ distribution towards more positive values is expected as the X-ray selection introduces an X-ray flux limit that biases us toward the X-ray brighter population at the optical luminosities probed with SDSS-V.
\hli{However, our X-ray selection does also identify X-ray weak populations as $\sim$ 7 per cent of the sample has $\Delta\alpha_\text{ox}<-0.2$. This figure demonstrates that we find both X-ray weak, X-ray normal,\footnote{We define `X-ray normal' here as $-0.2<\Delta\alpha_\text{ox}<0.2$.} and X-ray strong populations within our X-ray selected sample.} 
\hli{21/25 of the $\Delta\alpha_\text{ox}<-0.2$ sources are from \textit{Chandra}.}

Figure~\ref{fig:delta_ox} also presents the optically selected HiBALs studied by \citetalias{gibson_catalog_2009} in black, where they have also calculated their $\Delta \alpha_\text{ox}$ values relative to the \cite{just_2007} relation. The majority of the \citetalias{gibson_catalog_2009} HiBAL sample has $\Delta\alpha_\text{ox}<0$ \hli{although it does include a small number of objects with $\Delta\alpha_\text{ox} > 0$.}
However, the $\Delta\alpha_\text{ox}$ distribution of the \citetalias{gibson_catalog_2009} HiBALs does not align with the BI BAL quasar population isolated from our sample due to the differences in selection. Additionally, we observe that isolating the BAL quasars from the rest of the sample introduces a slight skew in our $\Delta\alpha_\text{ox}$ distribution towards lower values relative to the non-BALs, \hli{although the BALs still show a large fraction of $\Delta\alpha_\text{ox} > 0$ objects and many are X-ray stronger than those in the }\citetalias{gibson_catalog_2009} sample. In Table \ref{tab:KStest}, \hli{the \textit{p}-value for a KS test comparing the $\Delta \alpha_\text{ox}$ indicates that we should accept the hypothesis that the BI or AI-defined quasars and non-BALs are drawn from the same distribution.}  \citet{gibson_x-ray_2009} and \citetalias{wu_x-ray_2010} found that the absorption systems they labelled as mini-BAL quasars had X-ray properties more similar to the non-BALs than to the BALs. Our BI BALs and AI quasars do not appear to be X-ray weaker than the non-BALs and there is insufficient evidence to suggest that the X-ray-selected BALs preferentially show substantial X-ray weakness relative to non-BALs. \hli{While we do find X-ray weak objects, the lack of evidence of relative BAL X-ray weakness is likely due to the selection bias of our X-ray quasar sample that prefers X-ray strong quasars.} The $\Delta \alpha_\text{ox}$ values for our sample were also computed relative to the \citet{Timlin2021} relation, obtained for an optically-selected sample, finding the distributions of all the sub-populations agreed with Fig.~\ref{fig:delta_ox}.

\begin{figure*}
	\includegraphics[width=15.5cm,height=6.5cm]{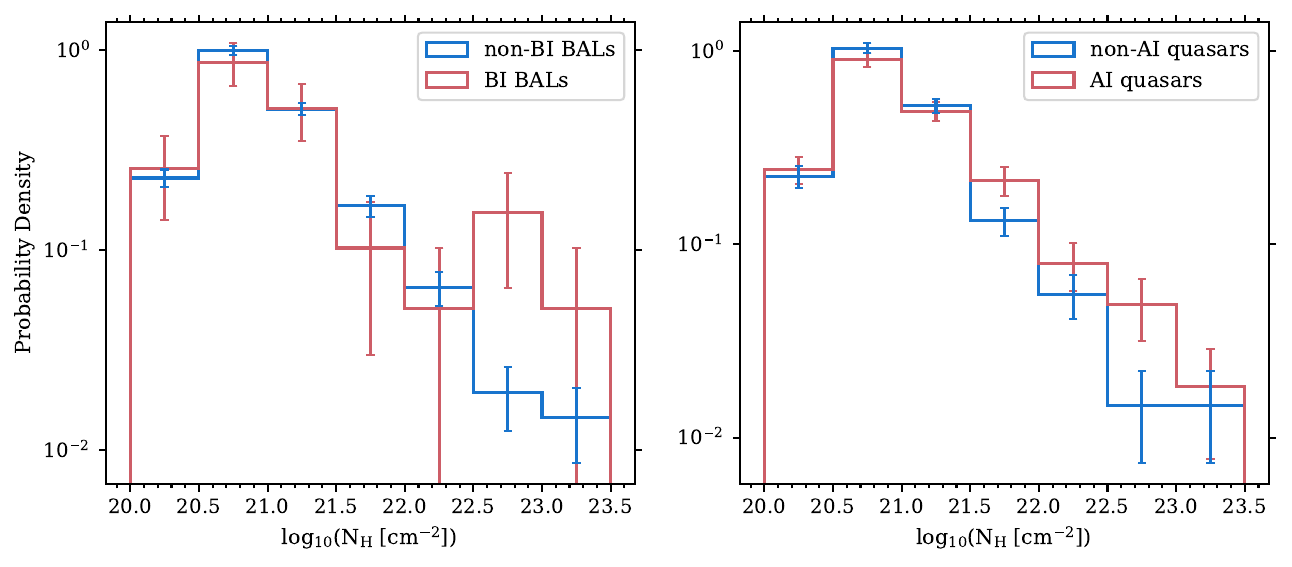}
    \caption{The distribution of column densities or $N_\text{H}$ (obtained from \citealt{liu_2022}) of the four sub-populations in the eFEDS sample where we have X-ray spectral information, with the lowest bin in each panel representing the unobscured fraction. While noting the overall small counts of the populations and the simplistic X-ray spectral model considered, the BAL and non-BAL quasars appear to be similarly distributed, i.e., the BAL quasars do not have higher column densities relative to the non-BAL quasars.} 
    \label{fig:Nh_plot}
\end{figure*}

\citet{liu_2022} performed X-ray spectral fitting of the eFEDS sample in order to derive various properties including the intrinsic column density, $N_\text{H}$, of the sources; however, note that the absorption model is likely too simplistic for BAL X-ray absorption. Figure~\ref{fig:Nh_plot} presents these column densities for our eFEDS sub-sample of quasars. We include in the figure those that are classified as `unobscured', `mildly-measured', and `well-measured' via the neutral absorption power-law model.\footnote{Model 1 of \citet{liu_2022}.} We do not use the tabulated $N_\text{H}$ values for the unobscured population; all of these objects are placed in the lowest $N_\text{H}$ bin (see section 4.3 of \citealt{liu_2022} for details). As shown in the figure, the $N_\text{H}$ values for the BI BAL and non-BAL quasars as well as the AI and non-AI quasars are similar while noting the marginally higher number of BAL quasars with higher $N_\text{H}$ values. \hli{The four BI BAL quasars with $N_H>10^{22.5}$\,cm$^{-2}$ have a median (mean) $\Delta\alpha_\text{ox}\simeq-0.091$ (-0.119) placing them in the weaker half of the X-ray normal population. Future larger samples with detailed X-ray spectral analysis are required to investigate further, but the $N_\text{H}>10^{22.5}$\,cm$^{-2}$ sample could be indicative of the absorbed X-ray weak BAL population observed in previous optically selected studies.}

\section{Discussion}\label{sec:discussion}

The spectroscopic follow-up of eROSITA, \textit{Chandra}, \textit{XMM} and \textit{Swift} sources with SDSS-V provides a large sample of X-ray selected quasars to investigate BAL quasars relative to non-BALs. This synergy has enabled, for the first time, to study a large sample of X-ray selected BAL quasars, i.e., 143 BI BALs and 954 AI quasars identified in this paper. Previous studies that investigated the X-ray-selected BAL quasars include \citet{Giustini2008} who studied 54 sources, \citet{streblyanska_2010} who studied 88 sources, and \cite{page_2017} who studied six sources. Our increase in sample size allows us to probe a broader dynamic range of BAL physical properties and explore trends with other signatures of outflowing gas (e.g., {\CIV} blueshift). We have also consistently selected a large sample of non-BAL quasars that provides a robust comparison to the BALs and their properties. In the following sections, we discuss the implications of the results obtained in this paper and contribute to the discussions of these previous studies. 

\subsection{X-ray selection excludes the highest outflow velocities in emission but spans the full range of absorption velocities}\label{sec:discussion_1}

As discussed in Section~\ref{sec:intro}, previous studies that investigated optically selected quasars have found BAL quasars to be X-ray weaker relative to non-BAL quasars.
Before comparing the BAL quasars with the non-BAL quasars in our study, let us first consider the impact of the X-ray selection on the properties of our sample relative to those of the optically selected comparison sample from SDSS-IV and what it could mean for the nature of the winds.

The {\CIV} emission blueshift is often considered as evidence for disc winds (e.g., \citealt{Gaskell1982, sulentic_2000,Leighly2004,Richards2002, Richards2011}; but see \citet{gaskell_line_2013, gaskell_case_2016} for an alternative view). The distribution of quasars within the {\CIV} space alongside its correlation with the \ion{He}{II} EW \citep[a tracer of the strength of the far-UV SED;][]{Leighly2004} point towards radiative line driving as being an important mechanism for driving strong winds \citep{Baskin_2013, Baskin2015-xk, rankine_2020, temple_testing_2023}. In detail, radiatively efficient winds occur when the outflowing gas can be propelled to high velocities by radiation pressure caused by the accretion disc's UV emission; both physical arguments and numerical simulations support the notion that radiation-driven outflows are feasible only when the gas surrounding the disc is not highly ionized since highly ionized gas would have a low concentration of ions that are capable of producing the observed UV line opacities (\citealt{Proga2007, matthews_disc_2023}; but see \citet{baskin_radiation_2014} for a different view). Through X-ray selection, we are missing the quasars with the highest {\CIV} emission blueshifts ($\gtrsim3500$\,{\kms}), which is consistent with the scenario where X-rays can overionize the gas resulting in lower velocity outflows. Therefore, through X-ray selection, we preferentially probe the high X-ray luminosity and low UV emission velocity part of the quasar population. \citetalias{rankine_2020} reported that the observed BAL fraction increases with increasing {\CIV} blueshift for a given {\CIV} EW (without correcting for selection effects) thus by consequence we may be missing a large fraction of the BAL population \citep[in agreement with][]{Giustini2008}. Since we are not probing the whole BAL population through our X-ray sample, i.e., there are no high {\CIV} blueshift quasars in this sample, any conclusions drawn can only be applied to the low--to--moderate {\CIV} blueshift BAL/non-BAL quasar populations.

Despite the effect of X-ray selection on the occupation of the {\CIV} emission space, in Section~\ref{sec:BAL identification} we found the overall fraction of BAL quasars \hli{(we find 6\% for our sample) is comparable to the observed fractions in optically selected samples, that typically find 10\% to 15\% }(e.g., \citealt{Tolea_2002}, \citealt{Hewett_2003}, \citealt{knigge2008}, \citetalias{gibson_catalog_2009}). 
\hli{While our observed BAL fraction is slightly lower than found in optically selected samples,} it remains a surprising result given that BAL quasars are generally thought to be X-ray weak relative to non-BALs (\citealt{green_1995}; \citealt{laor_1997}; \citealt{brandt_2000}; \citealt{gallagher_exploratory_2006}; \citetalias{gibson_catalog_2009}; \citealt{Luo2014}; \citealt{saccheo2023}). However, while the BAL fraction has been observed to increase with {\CIV} blueshifts, most quasars (and thus most BALs) are at lower {\CIV} blueshifts, and excluding the highest {\CIV} blueshifts via X-ray selection does not lead to a large change in the overall BAL fraction.

We also find that while X-ray selection preferentially identifies sources with relatively low-velocity disc winds (as traced by the {\CIV} emission line), the velocities of the \emph{absorbing} gas span the full range of BAL velocities seen in the optically selected BALs (see Fig.~\ref{fig:vmax_dist}). Furthermore, at a given {\CIV} distance (i.e., emission line velocity), the velocities of the absorbing gas for both X-ray and optically selected BALs are consistent (see Fig.~\ref{fig:vmax}). Thus, the presence of absorbing gas and its properties appear to be consistent, regardless of the selection wavelength.

\begin{figure}
	\includegraphics[width=\columnwidth]{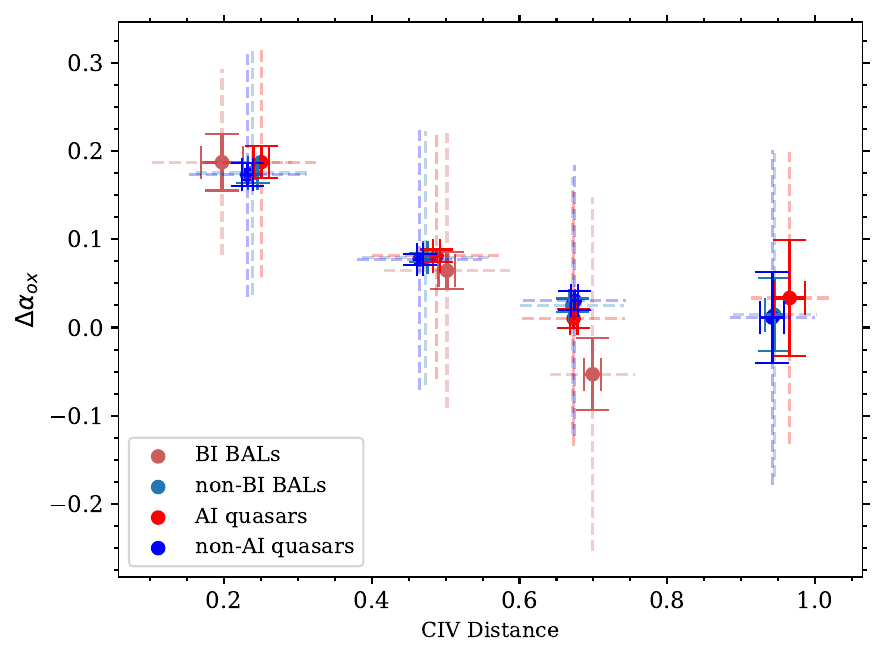}
    \caption{$\Delta \alpha_\text{ox}$ vs. {\CIV} distance for all the sub-populations in our X-ray selected sample, with each point showing the median $\Delta \alpha_\text{ox}$ value in different {\CIV} distance bins.
    While the rest of the sub-populations have four bins, the BI BAL quasars are not found in the last bin with the highest {\CIV} distance. 
    The error bars are similar to Fig.~\ref{fig:bi_vs_civ}. The trend suggests a decrease in median $\Delta \alpha_\text{ox}$ value with increasing {\CIV} distance for all but the highest {\CIV} distance bin, although the error bars are large -- quasars appear to be X-ray weaker at higher {\CIV} distances. 
    This trend is observed for all the sub-populations independent of their categorisation as a BAL or non-BAL. The trend suggests that, for a given quasar, low {\CIV} EWs and high {\CIV} blueshifts are associated with relatively X-ray-weaker systems.} 
    \label{fig:delta_ox_civdist}
\end{figure}

\begin{figure*}
	\includegraphics[width=16.5cm,height=10.5cm]
    {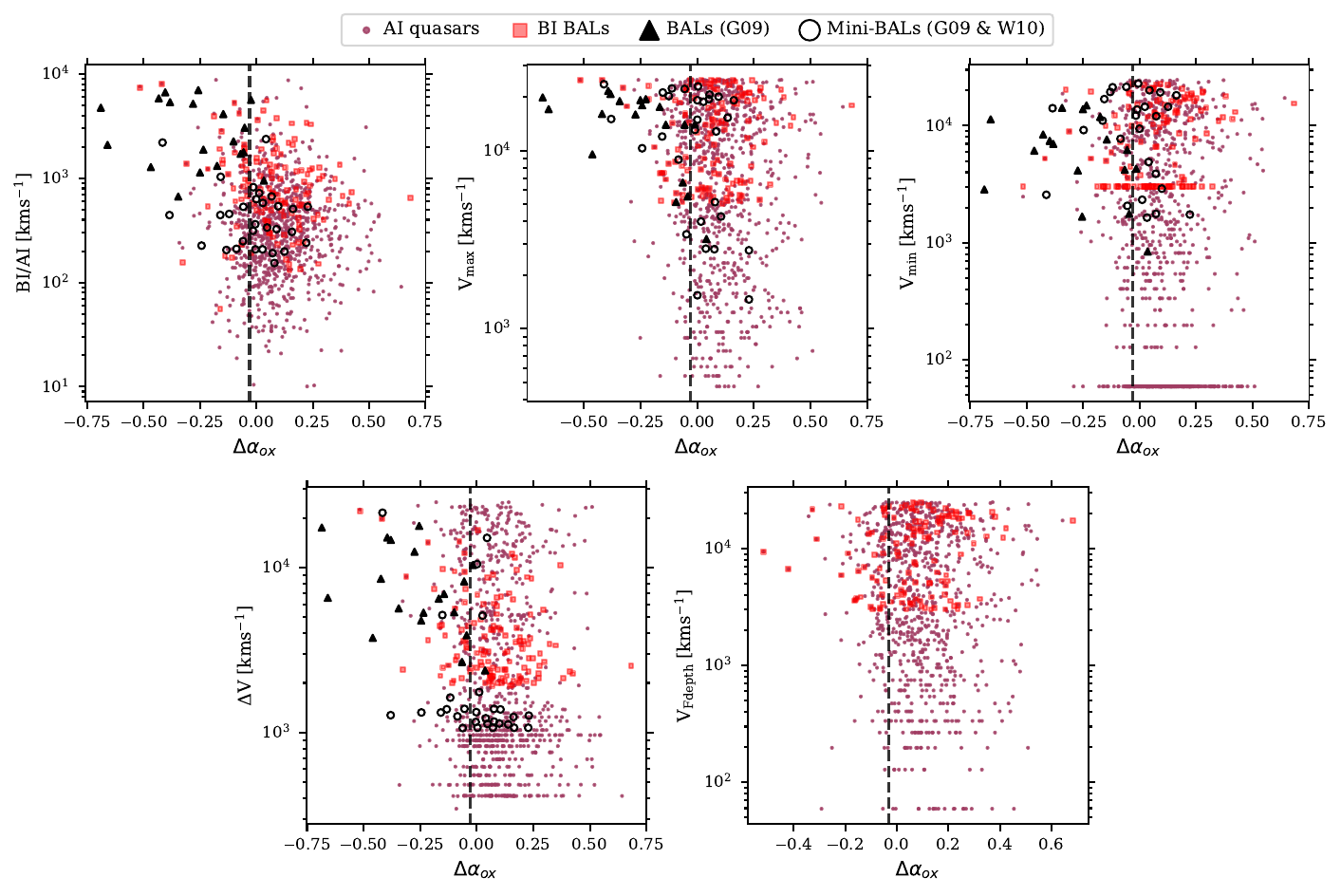}
    \caption{Various absorption properties as a function of $\Delta\alpha_\text{ox}$ for the BI and AI absorption systems. 
    \hli{The coloured dots (AI quasars) and squares (BI BALs) are from our sample. The black triangles are HiBALs from }\citetalias{gibson_catalog_2009} \hli{and open circles are mini-BALs from }\citetalias{gibson_catalog_2009} \& \citetalias{wu_x-ray_2010}, \hli{see figure 9 of }\citetalias{wu_x-ray_2010} \hli{for further details.}
    From top left: BI or AI value (as defined by Equations~\ref{eq:BI} and \ref{eq:AI}), maximum trough velocity, minimum trough velocity, velocity extent of absorption ($(\text{maximum}-\text{minimum})$ trough velocity which is the width of the trough for objects with one absorption feature), and velocity of deepest absorption.
    The velocity measurements show discretisation due to the finite width of the SDSS pixels. The build-up of objects at the limits of the BI and AI integration is most evident in the $V_\text{min}$ panel (top right). 
    The vertical dashed line in each panel shows the typical $\Delta\alpha_\text{ox}$ limit at the median X-ray sensitivity and median optical luminosity.
    \hli{The samples studied by} \citetalias{wu_x-ray_2010} \hli{fall below our $\Delta\alpha_\text{ox}$ limit and we do not find any clear trends within our X-ray flux limited sample, suggesting that these absorption properties do not depend on the X-ray properties for sources in our sample.} } 
    \label{fig:absprops}
\end{figure*}
The increase in BAL velocity with increasing {\CIV} distance (Fig.~\ref{fig:vmax}) is expected within a radiatively driven wind scenario since more line transitions within the gas lead to both stronger absorption features as well as the propulsion of the gas to higher velocities. However, the similar absorption velocities between the X-ray and optically selected samples indicate that the X-rays may have a relatively larger impact on the {\CIV} \textit{emission} line compared to the absorption. This behaviour is seen again in Fig.~\ref{fig:delta_ox_civdist} which shows the trend observed between $\Delta \alpha_\text{ox}$ and {\CIV} distance. The median $\Delta \alpha_\text{ox}$ value decreases with increasing {\CIV} distance for all the sub-populations, implying that systems with higher velocity {\CIV}-emitting outflows are X-ray weaker relative to those with lower velocity outflows. 
A similar anti-correlation between $\Delta \alpha_\text{ox}$ and {\CIV} distance was observed by \citet{timlin_correlations_2020} and \citet{Rivera_2022}. It was also an expected trend considering the relationship between \ion{He}{II} emission and {\CIV} blueshift (see Section~\ref{sec:intro} for references). \citet{Timlin2021} have also reported a strong correlation between $\alpha_\text{ox}$ and \ion{He}{II} across broad redshift and luminosity ranges.
Importantly for this paper, \citetalias{rankine_2020} found no difference between the BALs and non-BALs in regards to the \ion{He}{II} distributions in the {\CIV} space which aligns with the similar trends observed here between the X-ray selected BALs and non-BALs:
there is little difference in the $\Delta\alpha_\text{ox}$ trends with {\CIV} distance between the BAL and non-BAL quasars. 

\citetalias{wu_x-ray_2010} found correlations between various trough parameters and $\Delta\alpha_\text{ox}$ (see their figures 9--11) with slower and narrower absorption features present in objects with stronger X-ray emission. In Figure~\ref{fig:absprops}, we investigate how the absorption properties relate to the X-ray strength \hli{for our sample and compare with some of the results from figure 9 of} \citetalias{wu_x-ray_2010}. Unlike \citetalias{wu_x-ray_2010} we do not see any correlations for our X-ray selected sample. 
We caution, however, that the X-ray flux limit may be affecting any correlation by removing sources at low $\Delta\alpha_\text{ox}$. \hli{Many of the} \citetalias{wu_x-ray_2010} \hli{sample fall below our $\Delta\alpha_\text{ox}$ limit (see the vertical black dashed lines based on the median flux limit at the sample's median $L_{2500}$). However, our new sample illustrates the broad range of values that these measures of absorption properties span for our sample of sources at $\Delta\alpha_\text{ox} \gtrsim-0.1$, and that these properties do not appear to depend on the X-ray properties. This does not preclude a mild dependence for samples that span a broader range of $\Delta\alpha_\text{ox}$ but does suggest there must be a large scatter in any such relation.}
While we do not see a correlation between BI and $\Delta\alpha_\text{ox}$ (top-left panel), in Fig.~\ref{fig:bi_dist} we see that the X-ray selected nature of our sample is likely to have some effect on the types of BALs we are observing, since the BI distribution of the X-ray selected sample is missing the strongest BALs. However, apart from at the highest and lowest BI values, the BI distributions are very similar. The lack of a significant connection between the X-ray properties and the presence and properties of any absorption (either BI or AI troughs) furthers the conclusion that the wind signatures in emission are affected more than the wind signatures in absorption.
These results could be a manifestation of the different physical locations of the {\CIV}-emitting gas and the {\CIV}-absorbing gas. The emission line is produced from gas much closer to the X-ray corona (at sub-parsec scales) compared to the absorption that is thought to occur further from the central engine \citep[at hundreds to thousands of parsecs;][]{arav_evidence_2018, arav_hstcos_2020, xu_vltx-shooter_2019, miller_contribution_2020}. However, Extremely High Velocity Objects (EHVOs) are being found at distances closer to the broad emission line region (e.g., \citealt{rodriguez_hidalgo_extremely_2011, hamann_extreme-velocity_2013}; Rodr\'iguez Hidalgo et al. under review). Via the study of NALs, \citet{Bowler2014} and \citet{perrotta_hunting_2018} have shown that the ionization parameter (i.e., the number of ionizing photons per atom) of the outflowing gas decreases, and therefore distance from the quasar increases, with increasing absorber velocity (in BALs and non-BALs alike; \citealt{Bowler2014}). By this logic, the faster the BALs, the further out they may be where the impact of the X-ray emission is less pronounced, thereby reducing any correlation between the velocity of the wind and $\Delta\alpha_\text{ox}$. The AI population is likely also to be contaminated by intervening absorbers that will not be influenced by the X-ray properties of the background quasar.

\hli{To explore further any relationship (or lack thereof) between the velocity of the absorption systems and the X-ray properties,} we could extend this study to systems with even larger absorption speeds. EHVOs \citep{rodriguez_hidalgo_extremely_2011} extend the speed baseline as they are defined with minimum absorption speeds of 30,000 {\kms}. Most of these systems are found, due to the convenience of avoiding the Ly$\alpha$ forest, with maximum speeds of 60,000 {\kms}, although some exceptions with larger speeds have been found (\citealt{hamann_does_2018}, Hall et al. in prep). EHVOs also show, on average, larger speeds of {\CIV}-emitting outflows than BALs \citep{rodriguez_hidalgo_connection_2022}. Thus, studying the X-ray properties of EHVOs would allow us to test whether these systems are X-ray weaker relative to those with lower velocity outflows, both in emission and absorption. Testing the X-ray properties of EHVOs as a function of {\CIV} blueshifts would allow us to test whether X-rays have, as appears to be in the case of BALs, a larger impact on emission than absorption properties.

\subsection{BAL and non-BAL quasars span the same range of X-ray properties}\label{sec:discussion_2}

With X-ray information available for the BAL and non-BAL quasars in our sample, we have investigated the similarity in the properties derived from their X-ray data.
Previously studied optically selected samples have observed BAL quasars to have lower observed X-ray luminosities relative to non-BAL quasars, which is well-explained within the accretion disc wind model that invokes X-ray shielding (\citealt{Murray1995}; \citealt{proga_dynamics_2004}; \citealt{Luo2013}) to explain the X-ray weakness observed in BAL quasars. 

One key result of our paper is that we find BAL quasars across the full range of relative X-ray strength \hli{probed by our sample} (i.e., $\Delta\alpha_\text{ox}$; Fig.~\ref{fig:delta_ox}), and we do still identify moderately X-ray weak systems despite the bias due to the X-ray flux limits. \hli{Previous studies have also shown that HiBALs can be X-ray weaker than found in our sample} \citep[e.g., reaching $\Delta\alpha_\text{ox}\sim-0.7$:][]{gallagher_exploratory_2006} \hli{and thus BALs exist across the full range of relative X-ray strength of quasars}.
X-ray strong BALs are relatively rare and strong BAL features rare also. In an optically selected flux-limited sample one is biased towards optically brighter sources where X-ray strong objects are rarer still (see the lack of objects towards the top-right of Fig.~\ref{fig:alpha_ox}). In our X-ray selected sample, we are biased towards X-ray strong objects, particularly at low optical luminosities (see where the flux limit and $\alpha_\text{ox}$ relations intersect). Given the previous sample sizes of BALs with X-ray observations and the expected rarity of X-ray strong BALs, our sample selection is more likely to include these objects.

\begin{figure*}
	\includegraphics[width=15.5cm,height=6.5cm]{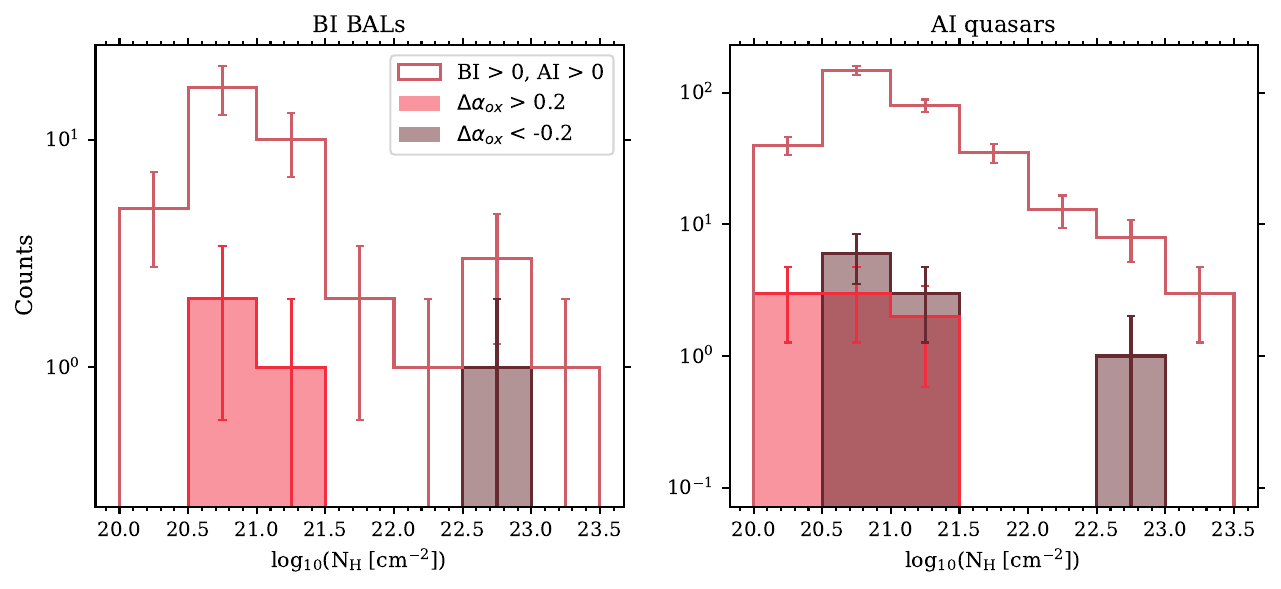}
    \caption{\hli{The distribution of column densities or $N_\text{H}$ of the BI BALs (left) and AI quasars (right) where we have X-ray spectral information, similarly as in} Fig.~\ref{fig:Nh_plot}. \hli{The open histograms contain the whole BI/AI samples (the same red histograms as in} Fig.~\ref{fig:Nh_plot}). \hli{The X-ray weak sub-populations are in brown and the X-ray strong in filled red. Noting the small sample sizes and the large error bars, one X-ray weak BI BAL has a large column density whilst the X-ray strong BI BALs have lower column densities. The majority of the X-ray strong and weak AI quasars, on the other hand, have $N_\text{H}\leq21.5$.}}
    \label{fig:Nh_plot_bals}
\end{figure*}

\hli{We raise two questions that others have also posed:} 1) Is the X-ray weakness in the BAL quasars that \textit{do} have $\Delta\alpha_\text{ox}<0$ a consequence of intrinsic X-ray weakness caused by a dim X-ray corona (relative to the disc) or due to large column densities of absorbing material along the line-of-sight? 2) Why in an X-ray strong source do we still see BALs if the X-ray emission is likely to over-ionize the wind?
We discuss these questions with the help of Fig.~\ref{fig:Nh_plot_bals} \hli{where we split the BI and AI quasars into X-ray weak ($\Delta\alpha_\text{ox}<-0.2$) and strong ($\Delta\alpha_\text{ox}>0.2$) and plot their $N_\text{H}$ distributions.\footnote{The sample is limited to the eFEDS sources as before.} While the sample sizes are small and the error bars large, all of the X-ray strong BI BALs have $N_\text{H}<10^{21.5}$\,cm$^{-2}$ and there is tentative evidence (one object) that the X-ray weak BI BALs are more likely to have larger column densities than the X-ray strong BI BALs. It should be noted that the X-ray normal BI BAL population contains objects up to $N_\text{H}\sim10^{23.5}$\,cm$^{-2}$. This result suggests that the X-ray weak BALs in our sample are X-ray weak due to large column densities along the line-of-sight and are not intrinsically X-ray weak. This absorption scenario is favoured in most cases} \citep[e.g.,][]{gallagher_exploratory_2006}.

\hli{Meanwhile, the majority of the X-ray weak AI quasars have $N_\text{H}<10^{21.5}$\,cm$^{-2}$, similar to the X-ray strong AI quasars. This result suggests that an enhanced absorbing column is not the cause of the X-ray weakness in the AI quasars.} 

\hli{When combining the X-ray weak, normal and strong eFEDS samples,} the $p$-values for the column densities and the redshift-matched observed (not corrected for obscuration) X-ray luminosities (see Table~\ref{tab:KStest}) suggest that the intrinsic X-ray properties of our BAL and non-BAL quasar samples are drawn from the same distribution. This analysis is presented for a spectral model with a neutral absorber; however, \citet{streblyanska_2010} find a more significant fraction of absorbed BAL quasars when X-ray spectral fitting was performed with an ionized absorber model. Nevertheless, we determined the intrinsic luminosity $p$-values for the eFEDS sub-samples (similarly as in Section \ref{sec:CIVspace} and using luminosities from \citealt{liu_2022}), finding the BI BALs/non-BALs are consistent with being drawn from the same underlying distribution ($p$ = 0.6) while the $p$-value (= 0.002) rejects this hypothesis for the quasars with and without AI-defined absorption systems.
This is in contrast to \citet{liu_2018} who found a higher percentage of HiBAL quasars that are \textit{intrinsically} X-ray weak (7--10 per cent) as compared to the percentage of non-BAL quasars \citep[$<2$ per cent;][]{gibson_are_2008}. \hli{We note, however, that eFEDS is biased towards less obscured objects and so the X-ray selection may be selecting against the obscured BAL quasars resulting in observed similarities between the X-ray properties of the BAL and non-BAL quasars that are not intrinsic to the whole population.} 

Of course, our line of sight may not be the same as the line of sight between the BAL gas and the inner corona, so geometry may play a role. 
\citet{Giustini2008} also observed a lack of X-ray weakness in a population of X-ray-selected BAL quasars, albeit a sample numbering only 54. They argued that the lack of X-ray weakness in their BALs can be accommodated within the theoretical model of the accretion disc wind by considering that the subset of BAL quasars that are X-ray strong (X-ray selected in their case) may be being viewed from smaller angles with respect to the accretion disc than the traditional optically selected BALs. We are therefore missing sources where our line-of-sight intersects with a large amount of inner shielding gas that absorbs X-rays (resulting in the observed X-ray weakness), suppresses the ionizing X-ray radiation and results in highest velocity UV outflows (seen in {\CIV} emission). However, as discussed in Section \ref{sec:intro}, \citet{baskin_radiation_2014} suggest RPC to address the problem of over-ionization without invoking an additional X-ray shielding gas: the radial compression of the absorbing gas by disc radiation prevents the gas from being over-ionized \hli{and thus BAL quasars are not required to be X-ray weak, but they can still exist.} Separately, \citet{hamann_extreme-velocity_2013} observed that mini-BALs approaching speeds of 0.1--0.2 $c$ do not have significantly higher column densities than their lower velocity counterparts. We do not find a difference in the X-ray luminosities (observed or intrinsic) of the BI BAL and non-BAL quasars or the AI and non-AI quasars identified in our paper, and the RPC framework could explain the efficient outflows we observe without requiring an additional shielding component.

\hli{Jets can also contribute to the observed X-ray luminosity} \citep[see reviews by][]{worrall_x-ray_2009, hardcastle_radio_2020}.
\hli{Since we are studying the X-ray selected population of SDSS-V, we do not remove quasars from our sample on the basis of their radio properties. Furthermore, our sample does not have uniform or complete radio coverage. Fourteen per cent (332 objects) of our sample is within the footprint of the 2nd data release of the LOFAR Two-Metre Sky Survey} \citep[LoTSS;][]{shimwell_lofar_2022} \hli{and 53 per cent (1239 objects) is within the Rapid ASKAP Continuum Survey footprint} \citep[RACS;][]{mcconnell_rapid_2020}. 
\hli{Regardless, X-ray emission from jets could be responsible for some of the X-ray strong objects in our sample.
Investigating the radio properties of the SDSS-V X-ray selected BAL quasars is beyond the scope of this paper; however, we state that out of the objects with radio coverage, 125 are radio-loud, all of which have $\Delta\alpha_\text{ox}>0$.}\footnote{\hli{We match to the optical counterpart catalogue of LOFAR by} \citet{hardcastle_lofar_2023}, \hli{and utilise a crossmatch between Legacy Survey DR10 and the RACS catalogue} \citep{hale_rapid_2021} \hli{performed using \textsc{nway}} \citep{salvato_erosita_2022}. \hli{We employ a radio-loud threshold of $\log R>2$ for LOFAR and $\log R>1.5$ for RACS sources where $R = L_\text{radio}/L_{2500}$. The thresholds are based on extrapolating} \hli{the traditional threshold of $\log R>1$ at 5\,GHz to 144\,MHz and 888\,MHz, respectively. Radio luminosities are calculated assuming synchrotron emission with $S_\nu\sim \nu^{-0.7}$.}} \hli{To our interest 1/16 ($\sim6$ per cent) BI BALs,} 
\hli{and 14/135 ($\sim10$ per cent) AI quasars with $\Delta\alpha_\text{ox}>0.2$ and within the radio coverage are radio-loud. However, radio-loudness criteria overpredict the fraction of AGN-dominated radio sources} \citep{yue_novel_2025} \hli{and so these fractions may be inflated. Nonetheless, by our criteria, jets are not sufficient to explain all of the X-ray strong BI BALs and AI quasars.}

Intrinsic X-ray variability may also be important to consider as the inner X-ray corona is likely to vary fairly rapidly \citep[hours to days;][]{mushotzky_x-ray_1993}; perhaps the X-ray strong phases ($\Delta\alpha_\text{ox}>0$) are too short-lived to have sufficient impact on the (further out) gas associated with the BALs.
In addition to intrinsic X-ray variability, studies by \citet{saez_long-term_2012} and \citet{timlin_long-timescale_2020} revealed that the long-term X-ray variability (rest frame 0.1--30 years) likely due to movement of any shielding gas is independent of the classification as a BAL or non-BAL (despite these BALs being X-ray weak) which could fit with the picture emerging here where the presence of a BAL is somewhat independent of the shielding gas. Regardless of timescale, any X-ray variability may be responsible for the broad $\Delta\alpha_\text{ox}$ distribution in the first place.

Moreover, we find that the median column densities of both BALs and non-BALs are constant with increasing {\CIV} distance suggesting the intrinsic X-ray properties do not change with {\CIV} distance, although it is important to note that smaller numbers available for the investigation of $N_{\text{H}}$, the simplistic X-ray spectral model, \hli{and the selection bias of the sample} used by \citet{liu_2022} may limit our ability to identify any changes. We also calculated Eddington ratios for our sample by determining black hole masses from the \ion{Mg}{II} line (see \citealt{Vestergaard2009}) to calculate $\textup{L}_{\textup{bol}}/\textup{L}_{\textup{Edd}}$, finding this to increase with increasing {\CIV} distance for all of the quasar populations as expected from optically selected studies (e.g., \citetalias{rankine_2020}; \citealt{temple_testing_2023}) and weak emission-line quasar studies \citep[e.g.,][]{luo_x-ray_2015}. \hli{See Appendix} \ref{appendix_extra} \hli{for figures showing these measurements of $N_\text{H}$ and Eddington ratio with {\CIV} distance.}

\citetalias{rankine_2020} showed that the BALs and non-BALs in the SDSS-IV (and precursors) optically selected sample are co-located in {\CIV} emission space with a changing BAL fraction across the space. Here we have investigated the X-ray selected sample from SDSS-V and find similar results with the KS tests not allowing us to rule out the possibility that the BALs and non-BALs have different {\CIV} blueshift-EW distributions. 
It is clear, however, that the X-ray properties of the BALs and non-BALs in our X-ray-selected sample are not statistically distinct from one another.

\section{Conclusions}\label{sec:conclusions}

This paper aimed to study the X-ray-selected SDSS-V BAL quasars within the redshift range of 1.5 $\le z \le$ 3.5. A sample of 2317 quasars with S/N $>$ 5 was selected based on best-fitting ICA spectral reconstructions and BAL quasars were identified using the BI metric. Other absorption systems were also identified via the AI metric. We have studied the UV outflow signatures in the BAL quasars and investigated their X-ray properties relative to non-BAL quasars. Our study of a large X-ray-selected sample builds on prior studies of BAL and non-BAL quasars that predominantly used optically selected samples, hence we have drawn comparisons with previous results from \citetalias{gibson_catalog_2009} and \citetalias{rankine_2020}. The main results of this paper are as follows:

\begin{enumerate}
    \item This X-ray-selected quasar sample consists of 143 or 6.2 per cent BI BAL quasars and 954 or 41.2 per cent quasars with AI absorption systems, i.e., the observed fractions are \hli{slightly lower than but } comparable to those previously found in optically selected samples. We note that these observed BAL fractions are different from the intrinsic fraction due to selection effects at play, \hli{including the X-ray flux limits, the optical-magnitude limit for SDSS-V spectroscopy, and the S/N required to identify BALs in the optical spectra.} 

    \item The X-ray selected quasars lie at lower {\CIV} blueshifts relative to optically selected quasars (see Fig.~\ref{fig:bal/nonbal}). However, within the X-ray selected sample, the BI BALs, AI quasars, and non-BAL quasars overlap in {\CIV} space. 
    While BALs span the same range of {\CIV} emission properties as X-ray selected non-BALs, we found small -- but statistically significant -- differences such that BALs are preferentially found in quasars with slightly lower {\CIV} EWs. 
    
    \item While the extent of the {\CIV} emission-line blueshifts differ for the optically and X-ray selected samples, the \emph{absorption} velocities are independent of the target selection (Fig. \ref{fig:vmax_dist} \& \ref{fig:vmax}), suggesting that X-rays may have a larger impact on the gas associated with {\CIV} \textit{emission} compared to the (likely distinct, physically further out) gas responsible for the absorption in BALs (see Section \ref{sec:discussion_1}).
    This picture is supported by the result that X-ray weaker sources are at higher {\CIV} distances (Fig.~\ref{fig:delta_ox_civdist}) and the lack of connection between the absorption properties of the X-ray selected BI BAL and AI quasars relative to X-ray strength or $\Delta\alpha_\text{ox}$ (Fig.~\ref{fig:absprops}).

    \item \hli{For the eFEDS sub-set of our sample with X-ray spectral fitting results,} 
    the observed X-ray luminosities (i.e., not corrected for obscuration) are the same for the BAL and non-BAL quasars (see Fig.~\ref{fig:xlum}). Additionally, we found a similar distribution of X-ray absorption column densities, $N_\text{H}$, (see Fig.~\ref{fig:Nh_plot}) suggesting that the \emph{intrinsic} X-ray properties of the BAL and non-BAL quasars studied in this paper are also similar -- validated for the eFEDS detected BI populations with a KS test $p$-value of 0.6 -- \hli{although we note that the soft X-ray eROSITA selection likely misses heavily obscured objects.}

    \item We found that our non-BAL quasars and the BI BAL and AI quasars overlap in the $\alpha_\text{ox}$ vs. $L_{2500}$ space (see Fig.~\ref{fig:alpha_ox}),
    and that \hli{BALs exist across the entire range of relative X-ray strength} (see Fig.~\ref{fig:delta_ox}) 
    \hli{probed with our X-ray selected sample. Due to this X-ray selection, our BAL and non-BAL subsamples are very similar, unlike the optically selected HiBALs studied by} \citetalias{gibson_catalog_2009} that show X-ray weakness.

    \item X-ray selection enabled the identification of examples of (relatively) X-ray bright BAL quasars, i.e., the  $\Delta\alpha_\text{ox} > 0$ range is more fully sampled through X-ray selection compared to the samples in \citetalias{gibson_catalog_2009} (see Fig.~\ref{fig:delta_ox}). \hli{Our sample spans a much wider range than }\citetalias{gibson_catalog_2009}, \hli{reaching $\Delta\alpha_\text{ox} > 0.4$, which includes sources that were not found by }\citetalias{gibson_catalog_2009}.
\end{enumerate}

Overall, the main result of this investigation is that the X-ray-selected BAL and non-BAL quasars studied in this paper appear to have similar X-ray properties, i.e., the X-ray properties of a given quasar are independent of its categorisation as a BAL or non-BAL. 
Further investigation of the intrinsic vs. observed X-ray properties of the X-ray selected BALs relative to the non-BAL quasars is required to understand the nature of X-ray weakness and brightness of these BAL quasars. X-ray spectral analysis using a larger sample and deeper X-ray data will be possible with future eROSITA data releases and will allow for further exploration of the column densities present in BAL quasars and the investigation into the presence (or absence) of any X-ray absorbing gas/shielding mechanism with greater statistics.

\section*{Acknowledgements}

PH, ALR, and JA acknowledge support from a UKRI Future Leaders Fellowship (grant code: MR/T020989/1). ALR also acknowledges support from a Leverhulme Early Career Fellowship. WNB acknowledges support from NSF grant AST-2407089, and the Penn State Eberly Endowment.
PRH acknowledges support from the National Science Foundation AAG Award AST-2107960 and the SDSS FAST-III program.
CA and ZI acknowledge the support of the Excellence Cluster ORIGINS, which is funded by the Deutsche Forschungsgemeinschaft (DFG, German Research Foundation) under Germany's Excellence Strategy – EXC-2094 – 390783311.
CR acknowledges support from Fondecyt Regular grant 1230345, ANID BASAL project FB210003 and the China-Chile joint research fund.
For the purpose of open access, the authors have applied a Creative Commons Attribution (CC BY) licence to any Author Accepted Manuscript version arising from this submission.

Funding for the Sloan Digital Sky Survey V has been provided by the Alfred P. Sloan Foundation, the Heising-Simons Foundation, the National Science Foundation, and the Participating Institutions. SDSS acknowledges support and resources from the Center for High-Performance Computing at the University of Utah. SDSS telescopes are located at Apache Point Observatory, funded by the Astrophysical Research Consortium and operated by New Mexico State University, and at Las Campanas Observatory, operated by the Carnegie Institution for Science. The SDSS web site is \url{www.sdss.org}.

SDSS is managed by the Astrophysical Research Consortium for the Participating Institutions of the SDSS Collaboration, including Caltech, The Carnegie Institution for Science, Chilean National Time Allocation Committee (CNTAC) ratified researchers, The Flatiron Institute, the Gotham Participation Group, Harvard University, Heidelberg University, The Johns Hopkins University, L’Ecole polytechnique f\'{e}d\'{e}rale de Lausanne (EPFL), Leibniz-Institut f\"{u}r Astrophysik Potsdam (AIP), Max-Planck-Institut f\"{u}r Astronomie (MPIA Heidelberg), Max-Planck-Institut f\"{u}r Extraterrestrische Physik (MPE), Nanjing University, National Astronomical Observatories of China (NAOC), New Mexico State University, The Ohio State University, Pennsylvania State University, Smithsonian Astrophysical Observatory, Space Telescope Science Institute (STScI), the Stellar Astrophysics Participation Group, Universidad Nacional Aut\'{o}noma de M\'{e}xico, University of Arizona, University of Colorado Boulder, University of Illinois at Urbana-Champaign, University of Toronto, University of Utah, University of Virginia, Yale University, and Yunnan University.

This work is based on data from eROSITA, the soft X-ray instrument aboard SRG, a joint Russian-German science mission supported by the Russian Space Agency (Roskosmos), in the interests of the Russian Academy of Sciences represented by its Space Research Institute (IKI), and the Deutsches Zentrum f\"{u}r Luft- und Raumfahrt (DLR). The SRG spacecraft was built by Lavochkin Association (NPOL) and its subcontractors, and is operated by NPOL with support from the Max Planck Institute for Extraterrestrial Physics (MPE). The development and construction of the eROSITA X-ray instrument was led by MPE, with contributions from the Dr. Karl Remeis Observatory Bamberg \& ECAP (FAU Erlangen-Nuernberg), the University of Hamburg Observatory, the Leibniz Institute for Astrophysics Potsdam (AIP), and the Institute for Astronomy and Astrophysics of the University of T\"{u}bingen, with the support of DLR and the Max Planck Society. The Argelander Institute for Astronomy of the University of Bonn and the Ludwig Maximilians Universit\"{a}t Munich also participated in the science preparation for eROSITA.

This research has also made use of data obtained from the Chandra Source Catalog, provided by the Chandra X-ray Center (CXC).

This research has made use of NASA's Astrophysics Data System and adstex (\url{https://github.com/yymao/adstex}).

\section*{Data Availability}

The SDSS-V data, including optical counterpart identifications to the X-ray sources as part of targeting and the optical spectra that underlie this article, are publicly available at \url{https://www.sdss.org/}. Data products including {\CIV} emission line measurements, BAL properties, and X-ray properties (excluding eRASS:1-derived properties) are available online (see Appendix~\ref{appendix_data}). The eRASS:1 optical counterpart identification will be made public in Salvato et al. (in preparation). After which, the eRASS:1 data will be made available upon reasonable request.
eROSITA data is available at \url{https://erosita.mpe.mpg.de}.
The parent \textit{Chandra} catalogue used for targeting is available at \url{https://doi.org/10.25574/csc2}.
The parent \textit{Swift}-XRT catalogue is available at \url{https://www.swift.ac.uk/2SXPS/exsess/} and the parent \textit{XMM-Newton} catalogue is provided at \url{http://xmm-catalog.irap.omp.eu/}.



\bibliographystyle{mnras}
\bibliography{main} 




\appendix

\section{Data} \label{appendix_data}
Table~\ref{tab:data_rows} contains the first 5 rows of the catalogue available on MNRAS. Table~\ref{tab:data_col} lists the descriptions of the available data columns. While the SDSS-V optical data and eRASS:1 X-ray data are separately in the public domain, the optical cross-match for this sub sample is not yet published. As such, we do not present the X-ray information for sources with ``eROSITA (eRASS:1)'' in the Instrument column of Table~\ref{tab:data_rows}. See the Data Availability section for more information.

\begin{table*}
\caption{The first 5 rows of the table containing optical and X-ray information for our sample. The full table is available online. See Table~\ref{tab:data_col} for a description of the columns.}
\label{tab:data_rows}
\begin{tabular}{ccccccccccc}
\hline
SDSS\_ID & RACAT & DECCAT & zfinal & Instrument & BI & BI\_err & BI\_VMAX & BI\_VMIN & BI\_VEXT & BI\_POSMIN \\
\hline
70345276 & 7.898 & 0.572 & 2.236 & Chandra & 5556.876 & 1.726 & 11719.405 & 3027.526 & 8691.879 & 6202.113 \\
70346150 & 8.987 & 1.280 & 1.732 & Chandra & 0.000 & 0.000 & 0.000 & 0.000 & 0.000 & 0.000 \\
70837514 & 351.894 & 0.376 & 1.491 & XMM & 0.000 & 0.000 & 0.000 & 0.000 & 0.000 & 0.000 \\
101611989 & 136.440 & -1.538 & 1.615 & eROSITA (eFEDS) & 0.000 & 0.000 & 0.000 & 0.000 & 0.000 & 0.000 \\
55777672 & 138.105 & 3.988 & 1.636 & eROSITA (eFEDS) & 229.533 & 2.440 & 5167.040 & 3027.526 & 2139.514 & 3027.526 \\
\hline
\end{tabular}

\begin{tabular}{cccccccccc}
\hline
AI & AI\_err & AI\_VMAX & AI\_VMIN & AI\_VEXT & AI\_POSMIN & CIV\_EW & CIV\_blue & ERO\_ID & eFEDS\_ERO\_ID \\
\hline
5906.696 & 3.537 & 20597.718 & 956.732 & 19640.986 & 6202.113 & 71.107 & 1204.664 & -- & -- \\
133.793 & 4.242 & 7168.049 & 6202.113 & 965.935 & 6892.081 & 48.667 & 740.697 & -- & -- \\
0.000 & 0.000 & 0.000 & 0.000 & 0.000 & 0.000 & 70.718 & 194.348 & -- &  --\\
0.000 & 0.000 & 0.000 & 0.000 & 0.000 & 0.000 & 63.500 & 221.314 & -- & 3211 \\
840.909 & 5.857 & 5167.040 & 197.408 & 4969.632 & 1854.099 & 57.143 & 1412.510 & -- & 10482 \\
\hline
\end{tabular}

\begin{tabular}{ccccccccc}
\hline
SWIFT\_ID & XMM\_ID & CSC21P\_ID & L0.2-2.3keV & L2keV & L2500 & alpha\_ox & Daox & eddratio \\
\hline
-- & -- & 2CXO\_J003135.5$+$003421 & 45.151 & 44.728 & 46.372 & -1.630 & 0.046 & 0.452 \\
-- & -- & 2CXO\_J003556.8$+$011649 & 44.721 & 44.298 & 45.372 & -1.411 & 0.125 & 0.322 \\
8796114954158265 & 4XMM J232734.6$+$002233 & 2CXO\_J232734.7$+$002234 & 44.863 & 44.812 & 46.007 & -1.458 & 0.167 & 0.134 \\
-- & -- & -- & 44.718 & 44.373 & 45.201 & -1.317 & 0.195 & 0.110 \\
-- & -- & -- & 44.587 & 44.242 & 46.064 & -1.699 & -0.066 & 0.231 \\
\hline
\end{tabular}

\end{table*}

\begin{table*}
\caption{Description of the column names in Table \ref{tab:data_rows}.} 
\label{tab:data_col}
\begin{tabular}{p{0.15\linewidth} p{0.8\linewidth}}
    \hline
    Column Name & Description \\
    \hline
    SDSS\_ID &  Unique identifier for targets in SDSS.\\
    RACAT, DECCAT & RA and DEC of the objects (degrees). \\
    zfinal & The redshift of the objects, see section \ref{sec:data} for how these values were obtained. \\
    Instrument & The instrument from which the X-ray information is obtained.\\
    BI, AI &  The Balnicity Index and Absorption Index determined using equations \ref{eq:BI} \& \ref{eq:AI} along with 'BI\_err' (the error in BI), 'BI\_VMAX' (maximum velocity at which {\CIV} absorption occurs), 'BI\_VMIN' (the minimum velocity), 'BI\_VEXT' (width of the absorption trough) , 'BI\_POSMIN' (velocity of the deepest absorption) and equivalently for AI. All have units of {\kms}. \\
    CIV\_EW & The {\CIV} emission line EW in angstroms. \\
    CIV\_blue & The {\CIV} emission line blueshift in \kms. \\
    ERO\_ID &  eRASS:1 source ID. \\
    eFEDS\_ERO\_ID &  eFEDS source ID. \\
    CSC21P\_ID & \textit{Chandra} source ID. 
    \\ 
    XMM\_ID & The IAU name associated with the \textit{4XMM} source.\\
    Swift\_ID & \textit{Swift} source ID (2SXPS\_ID).\\
    L0.2-2.3keV & Logarithm of rest frame 0.2--2.3\,keV X-ray luminosity [$\log_{10}(\text{erg\,s}^{-1})$]. \\
    L2keV & Logarithm of monochromatic X-ray luminosity at 2\,keV [$\log_{10}(\text{erg\,s}^{-1})$]. \\
    L2500 & Logarithm of monochromatic optical luminosity at 2500\,{\AA} [$\log_{10}(\text{erg\,s}^{-1})$]. \\
    alpha\_ox & The X-ray-to-optical spectral index, see equation \ref{eq:alpha_ox}. \\
    Daox & The $\Delta \alpha_\text{ox}$ determined using equation \ref{eq:daox}. \\
    eddratio & The Eddington ratio determined using black hole mass estimates from the \ion{Mg}{II} emission line. \\
    \hline
\end{tabular}
\end{table*}

\section{BI BAL and AI $>$ 0 representative spectra} \label{appendix_spectra}

\begin{figure*}
\centering
\noindent\makebox[\textwidth][c]{%
\setlength\tabcolsep{0pt} 
\begin{tabular}{cc}
  \includegraphics[width=9.5cm,height=12cm]{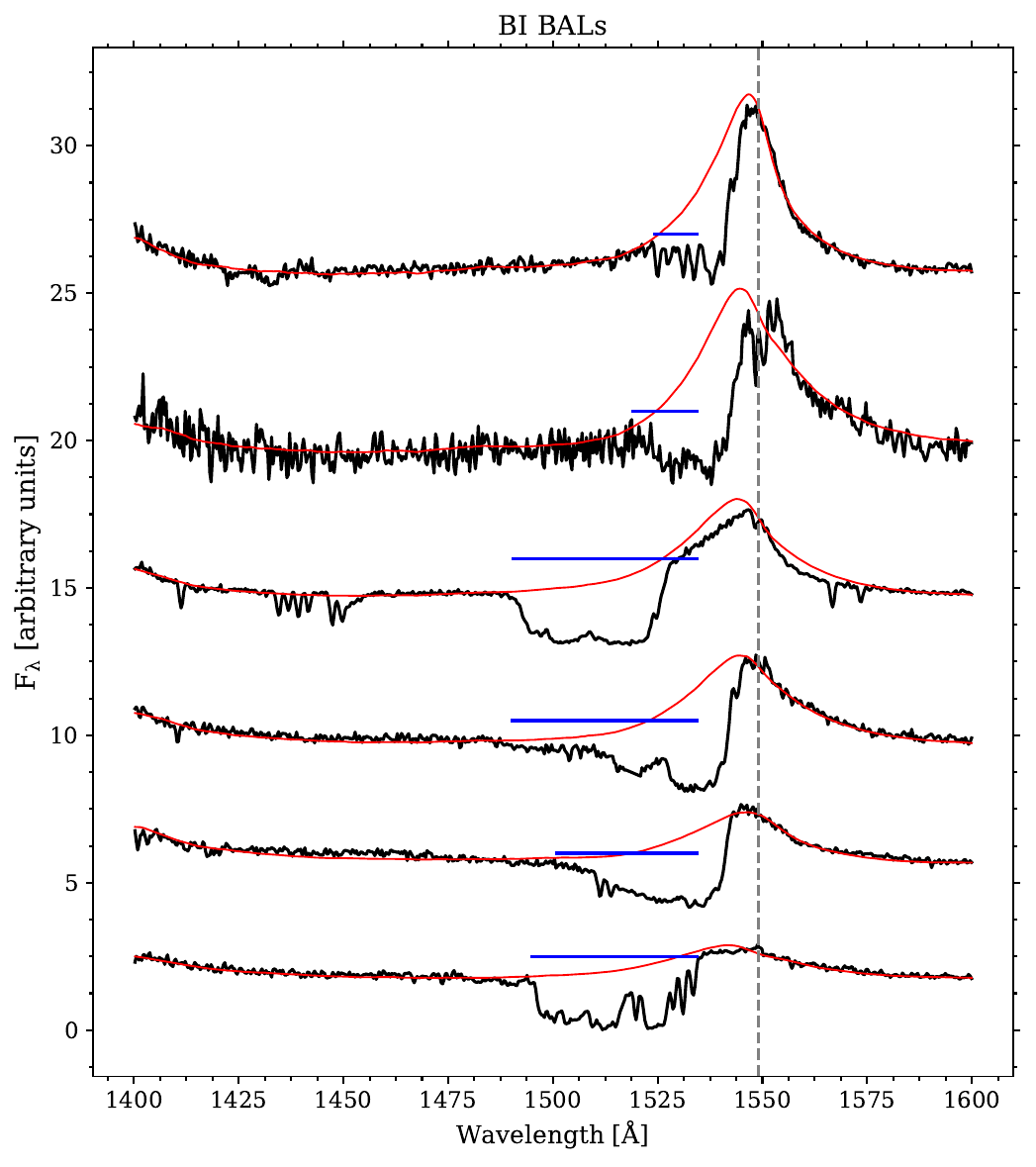} &\includegraphics[width=9.5cm,height=12cm]{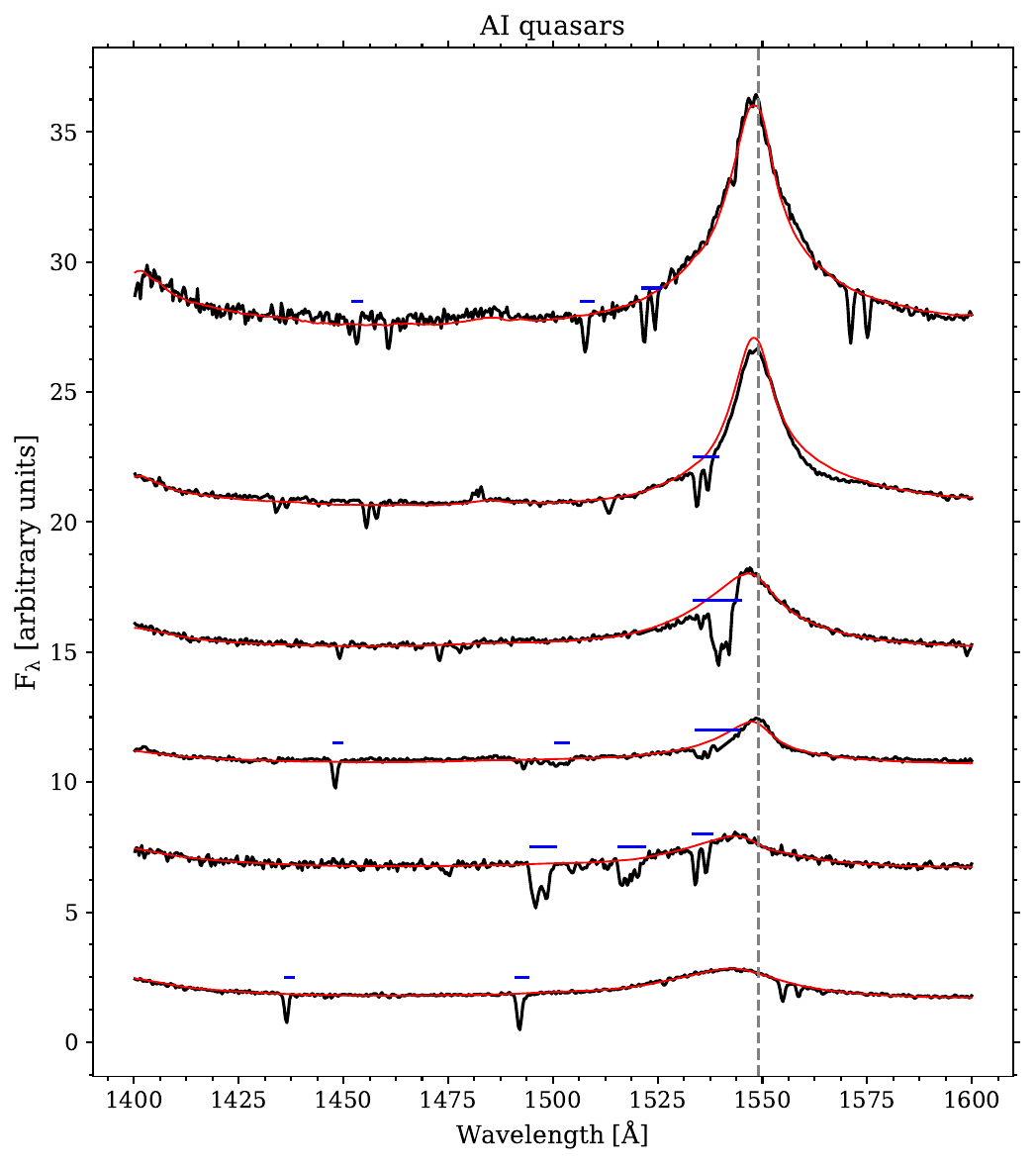} \\
\end{tabular}}
 \caption{Six representative spectra for the X-ray selected quasars identified as BI BALs (\textit{left panel}) and AI quasars (\textit{right panel}) in our sample. The rest-frame spectra are in black, their ICA reconstructions in red and the blue horizontal lines show where the absorption troughs were identified (see Equations \ref{eq:BI} \& \ref{eq:AI}). The dashed vertical line is at the {\CIV}$\lambda$1549 rest-frame wavelength, and the {\CIV} distance of the sources increases downwards in the panels. The BI BAL spectra show the expected broader and deeper absorption troughs as you move downwards in the plot. The AI quasar spectra show the presence of NALs and mini-BALs.}
 \label{fig:example_spec}
 \end{figure*} 

We present six representative spectra for the BI BALs and AI quasars identified in our X-ray-selected sample. Figure~\ref{fig:example_spec} shows the rest frame spectra in black and the reconstructions in red for quasars with BI- and AI-defined absorption systems using the definitions in Equations~\ref{eq:BI} and \ref{eq:AI}. The {\CIV} distance of the sources increases as we move vertically downwards and the troughs highlighted using horizontal blue lines indicate the absorption troughs classified and used to determine the BI or AI values ({\kms}) for these quasars. 

The left panel of Fig.~\ref{fig:example_spec} shows the representative spectra at different {\CIV} distances for the X-ray-selected BI BAL quasars. The horizontal blue lines in this panel highlight the 3000 {\kms} threshold set for the BI definition in Equation \ref{eq:BI} as many of the troughs extend to lower velocities which are not captured in the BI metric. As noted for the composites in Fig.~\ref{fig:composites}, the absorption troughs blueward of the {\CIV} line become wider and deeper with increasing {\CIV} distance, as expected and previously observed for the optically selected BALs studied by \citetalias{rankine_2020}.

The AI quasars in the right panel show the presence of NALs (as previously seen in composite spectra, see Fig.~\ref{fig:composites_ai}) as well as narrow troughs that may be associated with mini-BALs \citep[velocity widths of $\sim$1000--2000\,{\kms};][]{Trump2006}. For example, the middle trough of the fifth AI spectrum has a trough with a velocity width of 1173 {\kms}. Studies such as \citet{knigge2008} and \citetalias{rankine_2020} have previously noted the sub-populations that arise in the AI-defined quasar population. In detail, the AI definition overestimates the number of BAL quasars in the sample by the inclusion of NALs. The NALs have absorption troughs that are narrow enough ($\ge$450 {\kms}) to be included in the AI definition while excluded in the BI; therefore resulting in classically defined non-BALs with NAL signatures being categorised as AI quasars. As discussed in Section \ref{sec:CIVspace}, the NAL absorbers have very different internal kinematics relative to the BI-defined absorbers. 

The contrast in the absorption troughs in the left and right panels of Fig.~\ref{fig:example_spec}, suggests that the AI quasars are not representative of the `classical' BAL quasars with strong and broad absorption lines, as also noted by \citet{knigge2008}. We reiterate from previous studies that while the BI definition may lead to mild underestimation of BAL quasars, it is better at identifying characteristic BAL troughs and limiting contamination from other absorbers with different properties, e.g., NALs.

\section[N\_H and Eddington ratio with changing C IV distance]{$\bmath{N_\text{H}}$ and Eddington ratio with changing C\,{\sevensize IV} distance} \label{appendix_extra}

\begin{figure*}
\centering
\noindent\makebox[\textwidth][c]{%
\setlength\tabcolsep{0pt} 
\begin{tabular}{cc}
  \includegraphics[width=8.5cm,height=6.5cm]{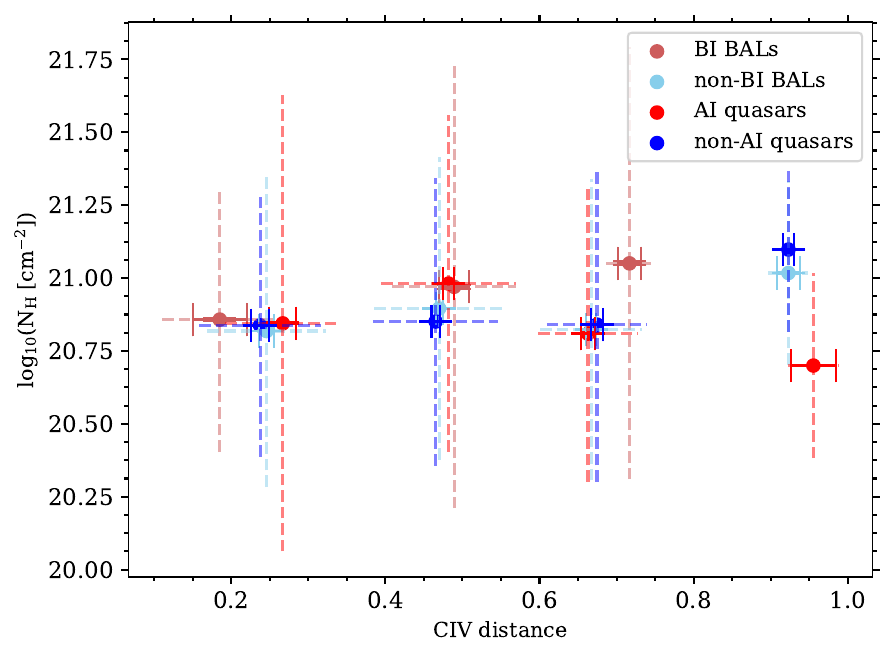} &\includegraphics[width=8.5cm,height=6.5cm]{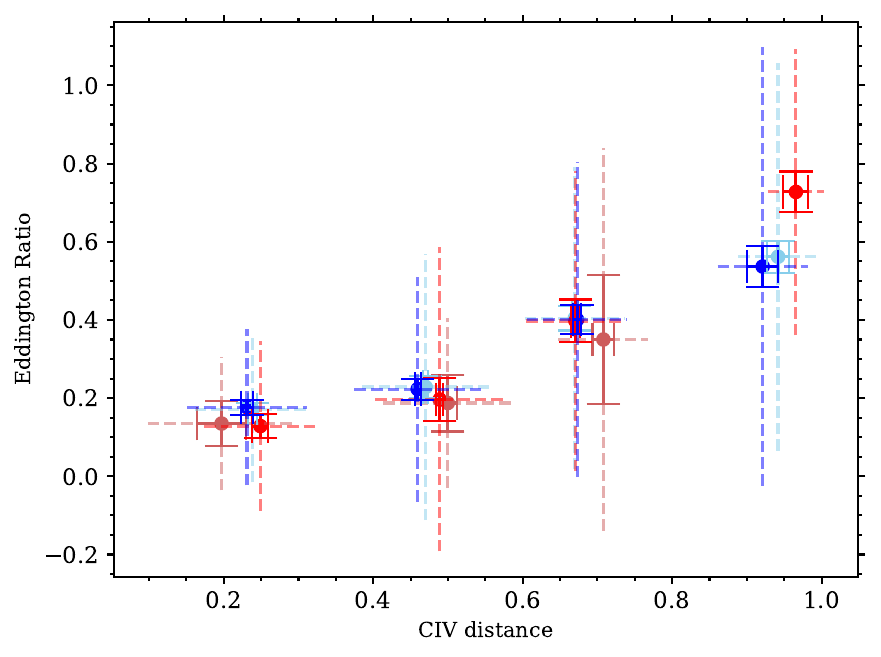} \\
\end{tabular}}
 \caption{$N_\text{H}$ vs. {\CIV} distance (\textit{left panel}) and the Eddington ratio vs. {\CIV} distance (\textit{right panel}) for all the sub-populations in our X-ray selected sample, with each point showing the median $N_\text{H}$ value/Eddington ratio in different {\CIV} distance bins. The dashed error bars indicate the standard deviation, and the capped error bars are the standard errors in the medians.
 As noted in Fig. \ref{fig:delta_ox_civdist}, the BI BALs are not found in the last bin with the highest {\CIV} distance. 
 For the smaller eFEDS sample in our paper we find that there is no trend for $N_\text{H}$ with {\CIV} distance, suggesting that the intrinsic X-ray properties do not change with {\CIV} distance. The Eddington ratio increases with {\CIV} distance as also previously observed in optically selected studies.
 These trends are observed for all the sub-populations independent of their categorisation.}
 \label{fig:appen_b}
 \end{figure*}

Figure \ref{fig:appen_b} shows the observed trends for $N_\text{H}$ (left panel) and Eddington ratio (right panel) with {\CIV} distance. While noting the caveats presented in the main text, we observe that $N_\text{H}$ does not change with the increasing {\CIV} distance. 

The Eddington ratios, calculated using black hole mass estimates via the \ion{Mg}{II} emission line, are observed to increase with {\CIV} distance. 
These trends in Figure \ref{fig:appen_b} are seen for all populations in our X-ray selected sample regardless of categorisation as BAL or non-BAL, AI quasar or otherwise.

\bsp	
\label{lastpage}
\end{document}